\documentclass[11pt]{article}
\usepackage{amssymb}
\usepackage{amsmath}
\usepackage{amsthm}
\usepackage[round]{natbib}
\usepackage{graphicx}
\usepackage{eurosym}
\usepackage{amsfonts}
\usepackage{multirow}
\usepackage{paralist}
\usepackage{verbatim}
\usepackage{subfig}
\usepackage{mathrsfs}
\usepackage{bm}
\usepackage{xcolor}
\usepackage{hyperref}
\usepackage{tabularx}
\usepackage{booktabs} 
\usepackage{endnotes}
\usepackage[linesnumbered,lined,boxed,commentsnumbered]{algorithm2e}
\usepackage{soul}
\usepackage{enumitem} 
\usepackage{mathtools}

\newtheorem{corollary}{Corollary}

\newtheorem{lemma}{Lemma}
\newtheorem{theorem}{Theorem}

\DeclareMathOperator*{\argmin}{arg\,min}
\DeclareMathOperator*{\argmax}{arg\,max}

\newtheorem{example}{Example}

\newtheorem{definition}{Definition}

\newtheorem{proposition}{Proposition}

\begin{document}
	
	\title{\LARGE\bf 
	Cooperative Agri-Food Export under Minimum Quantity Commitments}
	
	\author{Luis Guardiola$^{1}$\footnote{Email:  \url{luis.guardiola@ua.es}}, Behzad Hezarkhani$^{2}$\footnote{Corresponding Author, Email: \url{b.hezarkhanbi@soton.ac.uk}},   Ana Meca$^{3}$\footnote{Email: \url{ana.meca@umh.es}} \\
			$^{1}$	Universidad de Alicante, Departamento de Matem\'{a}ticas,\\
   03690 San Vicente del Raspeig, Alicante\\
   		$^{2}$ 
		Southampton Business School, University of Southampton,  \\		SO17 1BJ, United Kingdom,\\
   $^{3}$ Universidad Miguel Hern\'{a}ndez de Elche-I.U.\ Centro de Investigaci\'{o}n Operativa, \\
   03207 Elx, Alicante, Spain
		}
	\date{ }
	\maketitle

	\begin{abstract}
		International trade can be a profitable business for  agri-food communities.  However, access to international markets can be costly and thus unattainable for small and medium sized enterprises (SMEs). This problem is exacerbated under trade policies which require minimum quantity commitments (MQCs) on export volumes, e.g., licensing tariff rate quota (TRQ) mechanisms.
        We show how cooperative exporting among agri-food SMEs can tackle the barriers posed by the MQCs, and give  market access to a broader range of SMEs. We formulate a class of cooperative games associated with these situations and find a gain-sharing mechanism that result in allocations in their corresponding cores. Thus, grand coalitions of cooperative exporting SMEs can form in stable manners. 
        This allocation rule shares the export surplus only among the ``essential" SME exporters, that is, the players who are sufficiently cost efficient. Thus, less cost efficient ``complimentary'' SMEs whose capacities are needed to maintain MQCs receive no benefit from collaborative exporting and their participation have to be altruistic.  We propose two modifications to our original allocation rule to  share a portion of export surplus among the complementary SMEs through taxing the essential SMEs: the first through egalitarian, and the second through revenue-based rates. We compare the performance of these allocations with the numerical examples and discuss their practical implications.  \\

	\end{abstract}

  \newpage
\section{Introduction}
Trade brings about vital business opportunities for businesses in the food supply chain, while also helping to diminish global food insecurity and expanding consumer choices for various goods. The trade in agricultural and food products has experienced substantial growth in the past twenty years, averaging an annual increase of nearly 7\% in real terms from 2001 to 2019. Agri-food trade is not only on the rise but is also progressively evolving into a truly global phenomenon \citep{OECD:web}. Agri-food trade is also heavily reliant on logistics to run the supply chains with costs rallying up against  the farm-gate value \citep{hummels2007transportation}.

On the other hand, agricultural products have long held an exceptional place in international trade, with protectionist policies being the norm rather than the exception.  The sector is often described as remaining sheltered from the large reductions in tariffs, and it is only with the 1994 Uruguay Round Agreement that modest tariff cuts were agreed \citep{bureau2019agricultural}. A host of tariff and non-tariff barriers impede the trade of agri-food products.\footnote{Non-tariff barriers comprise all policy measures other than tariffs and tariff-rate quotas that have a more or less direct incidence on international trade. They can affect the price of traded products, the quantity traded, or both \citep{gourdon2020non}.} Tariff Rate Quota (TRQ) mechanisms are the dominant form of tariff barrier in agricultural trade where the destination market rations the volume of exports by charging prohibitive tariffs on exports above certain thresholds \citep{skully200723}. 

In this paper, our setting resembles the case of exporting under licensing TRQ mechanisms, which are the most common implementation method of TRQ mechanisms \citep{hranaiova2006economics}.  
A licensing mechanism operates in the following manner. Before the quota period begins, potential exporters are invited to apply for export licenses. Applicants specify their intended quantity of exports. After an exporter receives a license, he/she has the right to export their allocated license quantities at a lower tariff rate which would apply if export is done without a license. Deviating from the allocated license quantities involves penalties.   Many countries also specify a minimum license amount \citep{skully1987economics}. Such  a minimum quantity commitment  (MQC), the term we use in this paper, specifies the minimum license quantity that can be allocated to an exporter. Some examples are provided next.  
The Australian government sets MQCs for some commodities as follows: 1 tonne for EU high quality beef, 500 tonnes for cotton exported to India, 1 tonne for carrots and  potatoes exported to Indonesia, 1 tonne for FTA (Free Trade Agreement) butter and for FTA cheddar cheese for export to the United States of America, among others.\footnote{Export Control (Tariff Rate Quotas—General) Rules 2021; made under subsection 432(1) of the Export Control Act 2020 and item 92 of Schedule 3 to the Export Control (Consequential Amendments and Transitional Provisions) Act 2020.} 
The Australian government also indicates penalties for underused  quota entitlements, and that the amount of the allocation penalty is the amount of the person’s tariff rate quota entitlement for the quota type in the next quota year.\footnote{Export Control (Tariff Rate Quotas—General) Rules 2021; made under subsection 432(1) of the Export Control Act 2020 and item 92 of Schedule 3 to the Export Control (Consequential Amendments and Transitional Provisions) Act 2020.}
In Canada, license holders that are unable to use their licenses will have their allocations in the following year adjusted downward in proportion to the amount they did not use.\footnote{Government of Canada-Tariff rate quotas explained - Frequently Asked Questions \url{https://www.international.gc.ca/trade-commerce/consultations/TRQ-CT/dpe-lvo-questions.aspx?lang=eng})}
The UK government indicates that new sugar export applicants  must provide evidence to prove that they have exported at least 25 tonnes of sugar during the 12-month period ending two months before the first application.\footnote{Notice to Traders 40/22 - exports of Sugar under Statutory Instrument 2020 No.\ 1432.} 
 
 In the contemporary landscape of international trade, small and medium-sized enterprises (SMEs) often find themselves facing formidable challenges when competing against larger industry counterparts. 
 The inherent limitations of SMEs, such as restricted capital, limited production capacity, and narrower market access, can impede their ability to compete on a global scale \citep{fliess2006role}. This hinders the profitability of trade and deprives such enterprises from important growth opportunities.
  By joining forces through collaborative initiatives, SMEs can pool their resources, share expertise, and collectively navigate the complexities of international trade. This collaborative approach not only mitigates individual weaknesses but also creates synergies that enhance the overall competitiveness of the participating SMEs. 
There is a considerable amount of evidence and research on how  SMEs can facilitate their access to global markets through collaborative exporter groups, a.k.a.\ exporter consortia (e.g., \citet{forte2019role}).\footnote{The export consortia agreement signed by five Jordanian food supplements factories is a real-world example of successful collaboration supporting SMEs  in Jordan. This joint effort aims to boost the private sector's contribution to job creation and economic inclusion, particularly for youth and women. Despite the challenges of COVID-19, the collaboration focuses on leveraging the valuable knowledge within the local food supplement sector to explore export markets. This initiative, extended to sectors like garment and fashion, natural cosmetics, and food supplements  exemplifies a collective endeavor to access non-traditional markets. Notably, the agreement with professional uniform clothing manufacturers aims to create an integrated collection targeting east Africa, demonstrating the potential for collaborative ventures to open new opportunities for SMEs.}

Export consortia emerge as collaborative solutions, enabling SMEs to pool resources for joint export initiatives. Still, clear mechanisms for cost/gain sharing in such consortia have never been discussed. In this paper, we model cooperative games associated with agri-food export consortia, taking into account practical challenges such as MQCs, and examine cost/gain sharing mechanisms with desirable properties that could stabilize such organizations.

We construct a model wherein  SMEs  engage in the production of a certain agri-food commodity and, in line with a requirement of common market access mechanisms, are required to commit to the quantities they intend to export to a specific international market. The committed export quantity has to be at least as large as a predefined  MQC  that is established by regulatory authorities. Initially, we analyze the optimal order quantities for each exporting company based on their costs and profit margins, leading to a categorization of firms according to their export strategies. Our focus then shifts to the examination of cooperation among SMEs, specifically those whose individual capacities fall below the MQC, a.k.a., SMEs. We demonstrate that identifying the optimal coalition of firms for export  is an NP-Hard problem. 
The 
rationale for such cooperation becomes evident, as no single company or group has an incentive to abandon the grand coalition due to a profit distribution rule that satisfies all SMEs. Additionally, we explore two profit distribution rules to justly compensate complementary SMEs, those that individually find exporting unprofitable but do so for the benefit of the group. Finally, we provide several illustrative examples showcasing optimal benefits and orders for cooperating SMEs, applying and discussing various proposed profit distribution rules

This paper is organized into five sections. Section 2 provides a review of the related literature. Subsequently, Section 3 introduces definitions and notations in cooperative game theory. In Section 4, we present the individual export model, both with and without an MQC, examining the behavior of exporters and their optimal order size. Moving to Section 5, we analyze a model derived from scenarios where multiple SMEs, each with a MQC, collaborate to form a coalition. This section ranks the SMEs based on their costs and benefits, determining their optimal order size. In Section 6, we examine the cooperative model and demonstrate that identifying the subset of firms within a coalition that should engage in export activities to achieve the optimal profit is an NP-hard problem. Section 7 defines the cooperative game arising from this model, explores its primary properties, and establishes the stability (in the sense of the core) of the Non-essential Exporter Altruistic rule, distributing all profits among exporters without allocating any to others. Section 8 shifts focus to two alternative profit allocation methods that compensate complementary enterprises for their contribution to the overall profit increase. Finally, Section 9 offers a comprehensive summary of the paper's outcomes and contributions, concluding with suggestions for future research directions.

\section{Literature Review}
The literature on international trade in operations, logistics, and supply chain management is scarce but growing. In \cite{charoenwong2023trade}, the authors employ firm-level global supply chain data, transaction-level shipping container data, and policy uncertainty indexes from prominent media outlets to investigate the correlation between policy uncertainty and shifts in supply chain networks. \cite{fan2022global} provide  an overview of operations and supply chain management research incorporating the role of political economy in global trade. \cite{lam2022impact} investigate the impact of foreign competition on the product quality of domestic firms. \citet{dong2020impact} examine the contemporary research on global supply chain management and study the effect of tariffs on the configuration of the global supply chain networks. \cite{cohen2020designing} explores research opportunities on how global supply chain modeling can inform the way firms react to changes in relevant government policies for manufacturing and logistics, including tariffs, content requirements, taxes, and investment incentives. \cite{nagurney2019tariffs,nagurney2019strict,nagurney2019global} developed a spatial-price network equilibrium model of TRQ mechanisms to analyze the joint export quantity decisions, route selection, and equilibrium prices. This paper contributes to this line of work by introducing a pragmatic approach to cooperation among SME exporters to bypass challenges posed by MQC requirements in international trade.

There is a substantial body of work on trade mechanisms in the economics literature. Among the most notable contributions is \citet{skully1987economics} who develops  a basic static model for TRQs and investigates different administration methods, including FCFS  and license  on demand. In \cite{boughner2000economics}, the authors study the economic effects of TRQ mechanism following the Uruguay Round Agreement on Agriculture, focusing on changes in tariffs, quotas, and market conditions. They emphasize how export quota allocation and distribution procedures impact TRQ efficiency. \cite{gervais2003evaluating} examines the impact of discretionary and non-discretionary methods for allocating TRQs in the Canadian chicken industry. \cite{larue2007should} compares export tariffs and quotas in Canada's dairy industry, managed by a marketing board. The study explores welfare rankings between price-equivalent quotas and tariffs, considering different assumptions about the marketing board's powers. \citet{Hezarkhani2023} examines the logistical issues associated with TRQ mechanisms in international trade and show how the lead-time, warehousing, and the choice of logistics channel impact the performance of TRQ mechanisms in terms of fill-rate and policy-maker's revenue. Central to this paper is overcoming the constraints imposed by licensing TRQ mechanisms with regard to MQCs through cooperation.

This paper examines the horizontal collaboration among producers in international supply chains. It applies cooperative game theory to establish appropriate ways to share gains among cooperating enterprises in supply chains (see \citet{lozano2013cooperative} for a review of application domains in this context). Cooperative game theory has been used to address many supply chain problems, inasmuch as there are review papers tailored to specific contexts---e.g., cooperative transportation  \citep{cruijssen2007horizontal}, cooperative logistics network design \citep{hezarkhani2021collaboration}, cooperative inventory management \citep{fiestras2011cooperative,fiestras2012cost,fiestras2013new,fiestras2014centralized,fiestras2015cooperation}, cooperation in assembly systems \cite{bernstein2015cooperation}, cooperative sequencing \citep{curiel2002sequencing},  cooperative advertising \citep{jorgensen2014survey}, cooperative distribution chain \citep{guardiola2007cooperation} and \citep{GUARDIOLA2023102889} among others. Still, there are many ad hoc application domains, e.g., cooperative procurement \citet{hezarkhani2019s}, responsible sourcing \citep{fang2020cooperative}, vaccination supply chains \citep{Westerink-Duijzer2020}. In \cite{nagarajan2008game}, \cite{meca2008supply} and \cite{rzeczycki2022supply} we find different surveys that examine the applications of cooperative game theory in the context of supply chain management. Our work constitutes yet another area of supply chain management which can be fruitfully analyzed through cooperative game theory.

Operations research has been intertwined with cooperative game theory to develop dedicated methods.  \citet{anily2010cooperation} and \citet{karsten2015resource} study cooperation in service systems where enterprises pool capacities to serve costumers and studies the core of these games. \citet{armony2021pooling} further incorporates the strategic behaviour of customers in analyzing the pooling in service queues. \citet{he2012polymatroid} further examines the submodularity of objective functions in joint replenishment games.  \citet{liu2018simultaneous} introduces an instrument for maintaining the stability of grand coalitions by penalizing the deviations of the enterprises. \citet{fang2014stability} studies cooperative inventory transshipment games where endogenous coalition formation is allowed.
Our paper combines OR models with the concept of core to characterizes conditions for stable cooperation among SME exporters to international markets.

\section{Preliminaries cooperative game theory}

A cooperative game with transferable utility (TU-game) comprises a set of players $N=\{1,2,...,n\}$ and a characteristic function $v$, which associates each subset of $N$ with a real number. The subsets of $N$ are referred to as coalitions, denoted by $S$. Formally, the characteristic function is a mapping $v:2^{N}\longrightarrow \mathbb{R}$ such that $v(\emptyset)=0$. The value $v(S)$ of the characteristic function represents the maximum benefit that members of the coalition $S$ can achieve through cooperation. The coalition consisting of all agents, $N$, is termed the grand coalition. The subgame related to
coalition $S$, is the restriction of the mapping $v$ to the subcoalitions of $S$. 

One of the central inquiries in cooperative game theory revolves around the equitable distribution of benefits within the grand coalition after its formation. This distribution is achieved through allocations, represented by a vector $x\in \mathbb{R}^{n}$, where $n$ denotes the number of players in the set $N$. The class of superadditive games is particularly intriguing, serving as a motivation for the establishment of the grand coalition as it guarantees maximum profits for the coalition. Formally, a transferable utility (TU) game $(N,v)$ is deemed superadditive if, for every two coalitions $S,T\subseteq N$ with $S\cap T=\emptyset$, it holds that $v(S\cup T)\geq v(S)+v(T)$. Furthermore, TU-games characterized by higher profits for larger coalitions are termed strictly increasing monotone games. This can be expressed equivalently as $v(S)\leq v(T)$ for all $S\subseteq T\subseteq N$.  An imputation for the game $(N,v)$ is an allocation which verifies $\sum_{i\in
N}x_{i}=v(N)$ and $x_{i}\geq v(\{i\})$ for all $i\in N.$ The set of
imputations of the game is denoted by $I(N,v).$ 

Cooperative game theory provides diverse approaches for dividing the profits arising from collaboration. Two main categories of solutions exist: set solutions and point solutions. Set solutions involve the exclusion of allocations that fail to meet specific conditions, retaining only those that do. On the other hand, point solutions are derived through axiomatic characterization, meaning they are the unique allocations that satisfy specific properties.

The core of a  game is defined as the primary set solution. It encompasses all efficient allocations that maintain coalition stability, meaning no coalition has a motivation to exit the grand coalition without diminishing its overall profit. In formal terms,
\begin{equation*}
Core(N,v)=\left\{ x\in \mathbb{R} ^{n}:\sum_{i\in N}x_{i}=v(N)\text{ and} 
 \sum_{i\in S}x_{i}\geq v(S), \forall S\subset N \right\} .
\end{equation*}

The findings by \citet{bondareva1963some} and \citet{shapley1967balanced} offer a crucial criterion: the core of a  TU  game is guaranteed to be non-empty if and only if the game is balanced. This represents a key theorem in cooperative game theory.  It is a totally balanced game if the core of every subgame is nonempty. Totally balanced games were introduced in \citet{shapley1969market}. A game $(N,v)$ is regarded as convex if for all $i\in N$ and all $%
S,T\subseteq N$ such that $S\subseteq T\subset N$ with $i\in S,$ then $%
v(S)-v(S\setminus \{i\})\leq v(T)-v(T\setminus \{i\}).$ It is widely
acknowledged that convex games are superadditive, and superadditive games
are totally balanced. \citet{shapley1971cores} establishes that convex games are balanced and its core is large enough.


The nucleolus, a notable solution, is based on a notion of social equity. This solution
relies on the concept of excess. Given a cooperative game $(N,v)$ and an allocation $x\in \mathbb{R} ^{n}$, for any $S\subset N$ the excess of $S$ with respect to $x$ is defined by $e(S,x)=\sum_{i\in S}x_{i}-v(S)$. The excess function measures the degree of dissatisfaction of coalition $S$ with the selected allocation $x$. The coalition with the smallest excess is the coalition that most disagrees with the allocation. For any allocation $x\in \mathbb{R} ^{n}$, let $\theta (x)=\left( e(S,x)\right) _{S\subseteq N;S\neq \emptyset }\in \mathbb{R}^{2^{n}-1}$ be the vector of the excesses of the coalitions with respect to $x$ with the coordinates rearranged in decreasing order.  The nucleolus is the set of imputations which lexicographically maximizes the vector of excesses.
\begin{equation*}
\eta (N,v)=\left\{ x\in I(N,v):\theta (x)\succeq _{lex}\theta (y),\forall y\in I(N,v)\right\}
\end{equation*}

where $\theta (x)\succeq _{lex}\theta (y)$ means $\theta (x)=\theta (y)$ or
there exists $l,1\leq l\leq 2^{n}-1,$ such that $\theta _{k}(x)=\theta
_{k}(y)$ for all $k<l,$ and $\theta _{l}(x)>\theta _{l}(y).$ 
  Although the nucleolus is defined as subsets of allocations, \citet{schmeidler1969nucleolus} showed that each consists of a unique allocation. The nucleolus of a game whose core is non-empty belongs to the core and it is considered as the lexicographical centre of the core.



A point solution $\varphi$ refers to a function that, for each TU-game $%
(N,v) $, determines an allocation of $v(N)$. Formally, we have $%
\varphi:G^{N}\longrightarrow \mathbb{R}^{n}$, where $G^{N}$ denotes the
class of all TU-games with player set $N$, and $\varphi_{i}(N,v)$ represents
the profit assigned to player $i\in N$ in the game $v\in G^{N}$. Therefore, $%
\varphi(N,v)=(\varphi_{i}(N,v))_{i \in N}$ is a profit vector or allocation of $%
v(N)$. The nucleolus is a point solution. For a comprehensive overview of cooperative game theory, we recommend referring to \citet{gonzalez2010introductory}.

\section{Model}
A set of enterprises $N=\{1,...,n\}$ produce a common agri-food commodity. Each enterprise has the option to supply domestic market or export to international markets. For an arbitrary enterprise $i\in N$, we denote the production capacity with $Q_i>0$ which indicates the maximum amount of produce that the enterprise can supply. We normalize the cost of production  to zero for all enterprises.

Enterprises must decide how much to export.\footnote{This decision precedes the actual export time because of the production lead-time which could involve configurating the produce for the target market. Although market prices might be uncertain, we assume risk neutrality and consider the expected values.}
But prior to that, an enterprise wishing to engage in exports must commit a specific quantity of export  to the export control authority. This can be in the form of acquiring export licenses under a TRQ mechanism.\footnote{As we assume a penalty for over-supply, in theory, an exporter can commit to nothing and later on export and pay the corresponding penalty. However, this is not practical and as such we exclude this possibility.} The export commitment level of enterprise $i$ is denoted with $m_i \geq 0$. That is,  the export volume committed by the enterprise and approved  by the export control authority.
After making a commitment and at the time of export, an enterprise $i$ needs to decide the export quantity which is denoted by $0\leq q_i\leq Q_i$.

For every unit of exported produce, enterprises gain the export revenue $p\geq 0$.
 W.l.o.g, we normalize the revenue for supplying to domestic market to zero.  To be able to export, however, the enterprise $i$ has to incur a fix cost $c_i\geq 0$ in order to prepare the produce for international market.  For instance, this pertains to measures specific to production or those related to sanitary and phytosanitary requirements mandated by the target market.

Deviating from a committed export volume  entails two types of penalties: per unit over-supply penalty $r^o\geq  0$,  and per unit under-supply penalty $r^u\geq  0$. 
For example, in case the commodity is subject to a TRQ mechanism, if an enterprise exports above the volume of licenses that it posses, an over-quota tariff rate will be applied to every unit of excess export. Under-supply penalty can be applied per unit or be in terms of sanctions for future exports. For example, many governments curtail the future access to licenses for an enterprise who fails to fulfil its committed volume of exports.

It is a customary practice for authorities to establish a minimum threshold for the export commitment level (see the introduction for specific examples). We call this a minimum quantity commitment (henceforth MQC) and denote it with $\underline{m}\geq 0$. Thus, in order for an enterprise $i$ to export, it must make a commitment at least as large as the MQC, that is, $m_i\geq \underline{m}$.
The presence of positive MQCs has significant impacts on export decisions as we examine next.

\section{Individual Exporting}
We initiate our analysis by considering the non-cooperative scenario in which enterprises export individually. Let $\Delta_i:=Q_i p - c_i$ be the full capacity margin of exporting for an enterprise $i\in N$.  In order to examine the exporters decision making problems, we first consider the case with zero MQC, i.e., $ \underline{m}=0$ and later on incorporate $ \underline{m}>0$ into  the model.


\subsection{Exporting under zero MQC}
We refer to the case of  $\underline{m}=0$ as a setting without a (positive) MQC. In this case, an enterprise's choice of commitment has no positive lower bound, that is, $m_i\geq 0$. 
We formulate the individual profit function of the enterprises next.
The case of committing to zero exports ($m_i=0$) and the subsequent decision to  export nothing ($q_i=0$), leading to the normalized profit of zero, is used as the  benchmark against any positive export strategy. 
 For an enterprise $i\in N$, the choice of commitment level $ m_i>0$, and export quantity $0\leq q_i\leq Q_i$  results in the following profit:
\begin{equation}
     \Pi_i(q_i,m_i) =  
      \mathbf{1}_{q_i>0}   \Delta_i  - \left(q_i-m_i\right)^+ r^o -\left(m_i-q_i\right)^+ r^u  .
\end{equation}
where $\mathbf{1}_{q_i>0}=1$ if $q_i>0$ and $\mathbf{1}_{q_i>0}=0$ if $q_i=0$, and $(\cdot)^+=\max\{\cdot,0\}$. The optimization problem for $i$ is thus:
\begin{align}
    \max \quad & \Pi_i(q_i,m_i) \quad  s.t.\quad    m_i\geq 0,  0 \leq q_i\leq Q_i.     
\end{align} 
 The first observation is that without a positive MQC, agents bifurcate into two categories: those who fully utilize their capacity to export on the international market and those who focus on the domestic market. The Appendix provides the proofs of all results.

 \begin{lemma}
 \label{th1}
     Without a MQC required for export, i.e., $\underline{m}=0$, for every enterprise $i\in N$ the following holds:
     \begin{description}
         \item[(A) Exporters:] If  $\Delta_i\geq 0$ we have $q^*_i=m^*_i=Q_i$.
         \item[(B) Domestic producers:] Otherwise,  we have $q^*_i=m^*_i=0$. 
      \end{description}  
 \end{lemma}

\begin{figure}
\centering
\includegraphics[width=0.5\linewidth]{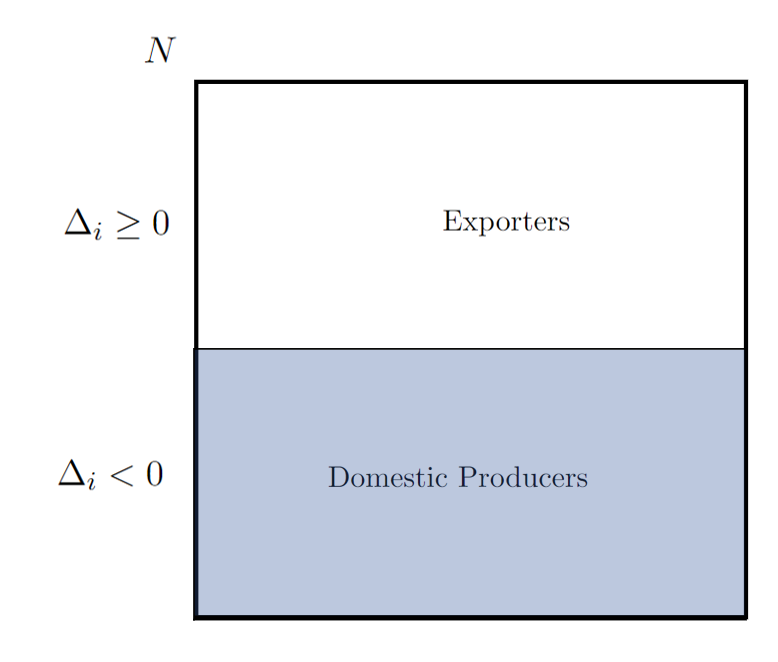}
\caption{Individual Exporting without MQC}
\label{fig:NoCom}
\end{figure}

Without a MQC required for export, every enterprise $i\in N$ with $\Delta_i\geq 0$ could commit his entire capacity for export, i.e., $m_i=Q_i$. Otherwise, any enterprise $i\in N$ such that $\Delta_i<0$ is better off supplying the domestic market only.  Figure \ref{fig:NoCom} demonstrates these relationship graphically.  
The optimal profit of the enterprise $i$ is the maximum of zero and his full capacity margin, that is:
\begin{equation}
    \Pi_i(q^*_i,m^*_i)=
    (\Delta_i)^+.
\end{equation}

\subsection{Exporting under positive MQCs}
We now incorporate the positive MQC constraint into the model. 
Define $\delta_i:=Q_i(p+r^u)-c_i$ as the under-supply adjusted (full capacity) margin of exporting for $i$.
With the introduction of $\underline{m}>0$, the optimization problem of an enterprise $i\in N$ becomes as
\begin{align}
\max \quad    \Pi_i(q_i,m_i) \quad s.t.\quad     m_i\geq \mathbf{1}_{q_i>0} \underline{m}, 0\leq q_i\leq Q_i.       
\end{align} 
The maximization problem must maintain a commitment at least as large as the MQC whenever a positive quantity is to be exported.
We follow the same logic as previous case to optimize the decisions on export quantity and commitment level.

  \begin{lemma}
 \label{th2}
     Given a MQC, $\underline{m}>0$, for any enterprise $i\in N$ the following holds:
     \begin{description}
         \item[(A) Type $\alpha$ Exporters:] If $Q_i\geq \underline{m}$ and $\Delta_i\geq 0$, we have $q^*_i=m^*_i=Q_i$.
         \item[(B) Type $\beta$ (SME) Exporters:] If $Q_i< \underline{m}$ and $\delta_i-\underline{m}r^u\geq 0 $, we have $q^*_i=Q_i$, and $m^*_i=\underline{m}$.
         \item[(C) Domestic producers:] Otherwise,  we have $q^*_i=m^*_i=0$.
     \end{description}    
 \end{lemma}

\begin{figure}
\centering
\includegraphics[width=0.5\linewidth]{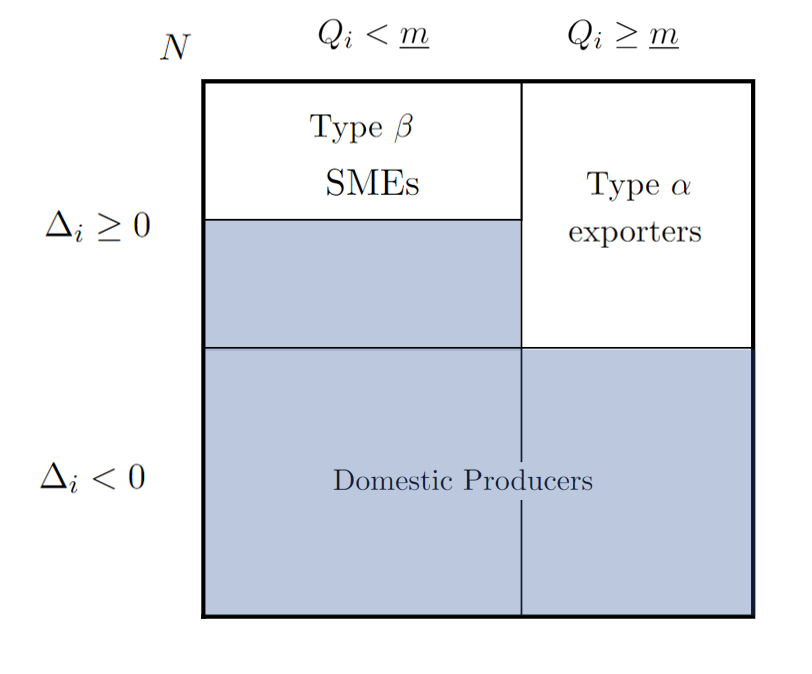}
\caption{Individual Exporting under MQC}
\label{fig:withCom}
\end{figure}

All exporting enterprises would utilize their entire capacity to  export. Type $\beta$ exporters, which corresponds to SMEs due to their capacities being smaller than $\underline{m}$, would have to under-supply with regard to the MQC and thus have to pay the under-supply penalty. Yet, type $\beta$ SMEs  find it profitable to do so.  Figure \ref{fig:withCom} demonstrates these relationship graphically.
 The profit function of an enterprise $i$ at optimality in this case is
\begin{equation}
    \Pi_i(q^*_i,m^*_i)=\begin{cases}
     \left( \Delta_i \right)^+    & \text{ if }  Q_i\geq \underline{m}  \\
    \left( \delta_i -\underline{m}r^u \right)^+ & \text{ if }  Q_i< \underline{m} 
     \end{cases}
 .
\end{equation}

Comparing to the case without MQCs, we can see that in this scenario SMEs, i.e., those $i$ with $Q_i<\underline{m}$, would only export if they can afford the under-supply penalty. 
Therefore, the capacity of an enterprise in comparison with to the MQC is a major factor that determines SMEs' export strategies.

\section{Cooperative Exporting}
As seen above, the introduction of MQCs makes it more difficult for SMEs to export. This can be remedied through collaboration with other SMEs. 
We develop the decision making problem for a group of SMEs $S\subseteq N$ with $Q_i<\underline{m}$ for all $i\in S$, that is, a coalition of SMEs.  From now on, we will refer to the enterprises as players. 

We define a collaborative exporting situation, abbreviated as a CE-situation,
by a tuple $(N,Q,C,p,\underline{m})$ where $N$
stands for the set of players, $Q=\left( Q_{i}\right) _{i\in N}$ denotes the
vector of production capacities (with $Q_{i}<\underline{m}$ for all $%
i\in N$), $C=\left( c_{i}\right) _{i\in N}$ represents the vector of fixed costs to export, and $p$ and $\underline{m}$ referring to the export margin and the MQC as defined previously, respectively.

Let $q=(q_i)_{i\in N}$ be the vector of export quantities (note that we do not restrict the vector to players in $S$ in order to simplify notation). Also, suppose that the commitment level of the coalition is fixed to $m\geq \underline{m}$. The profit function for  $S $ is:
\begin{equation}
    \Pi^S(q,m) =  
         \sum_{i\in S}\mathbf{1}_{q_i>0} 
         \Delta_i  	   - \left(\sum_{i\in S}q_i-m\right)^+ r^o -\left(m-\sum_{i\in S}q_i\right)^+ r^u 
\end{equation}
The optimization problem for the coalition of players in $S\subseteq N$ is 
\begin{align}
\label{coalprof}
    \max \quad  \Pi^S(q,m) \quad  s.t.\quad     m\geq  \mathbf{1}_{\sum_{i\in S}q_i>0} \underline{m}, 0 \leq q_i\leq Q_i, i\in S. 
\end{align}
Let $(q^S,m^S)$ be the optimal solution for $S$, i.e., the solution to the above problem. It follows immediately that $\Pi^S(q^S,m^S)\geq 0$. Given the structure of this optimization problem, it can be inferred that   over-supply is never occurs at optimality. Thus $r^o$ never impacts the optimal solution.


In order to develop observations about the nature of optimal export consortia (coalitions), we define three player types.

\begin{definition}
    Let $S\subseteq N$ be a coalition of player. We define the following terms:
    \begin{description}
         \item[Essential  SMEs of $S$:] $ S^{E}=\{i\in S|Q_i< \underline{m}, \Delta_i\geq 0 \}$ 
         \item[Potential  SMEs of $S$:] $S^{P}= \{i\in S|Q_i< \underline{m}, \delta_i\geq 0 \}$
         \item[Complementary  SMEs of $S$:] $S^{C}=S^{P}\setminus S^{E}.$
    \end{description}
\end{definition}

\begin{figure}
\centering
\includegraphics[width=0.5\linewidth]{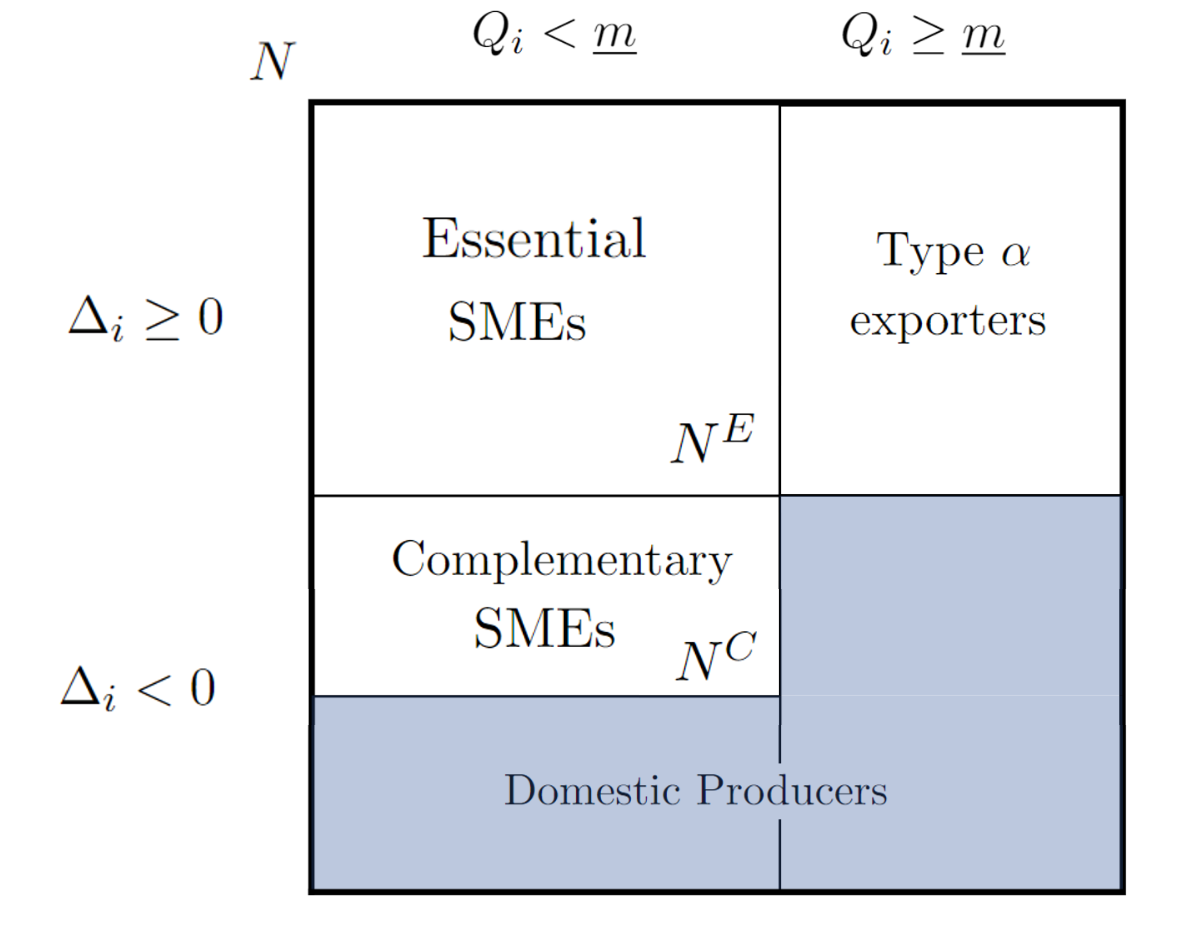}
\caption{Collaborative exporting under MQC}
\label{fig:CollwithCom}
\end{figure}
 
The \textit{essential} SMEs have non-negative full capacity margins. 
If an essential SME would not export on his own,  it would have been because of his capacity limitation rather than cost-inefficiency.
The \textit{potential} SMEs have non-negative under-supply adjusted  margins.  
Note that $S^{E}\subseteq  S^{P}$ since non-negativity of full capacity margins implies the non-negativity of under-supply adjusted  margins. 
The \textit{complementary} SMEs are potential but not essential. 
For every players in $S^{C}$ we have $\Delta_i< 0$ and $\delta_i\geq 0$. 
Figure \ref{fig:CollwithCom} demonstrates the different types of players in cooperative exporting graphically.

To focus on SMEs, from this point on we assume that $N$ is solely comprised of SMEs, that is, we assume for every $i\in N$ we have $Q_i<\underline{m}$.
The following theorem illustrates the optimal export strategies (quantities and commitment levels) of a coalition and shed light on the the role of different types of players. 
\begin{theorem}
\label{th3}
Consider a coalition $S\subseteq N$ and let  $(q^{S},m^S)$ be an optimal solution for $S$. If $\sum_{i\in S}q^S_i>0$, then we have the following cases:
\begin{description}
\item[(A)] If $\sum_{i\in S^{E}}Q_{i}\geq  \underline{m}$, then $m^{S}=\sum_{i\in S%
^{E}}Q_{i}$,   $q_{i}^{S}=Q_{i}$ for all $i\in S^{E}$,
and $q_{i}^{S}=0$ for $i\in S\setminus S^{E}$.

\item[(B)] If $\sum_{i\in S^{E}}Q_i< \underline{m}$, then 
for some $R^{S}\subseteq S$ such that $S^{E}\subseteq R^{S}\subseteq S^{P}$, we have $m^{S}=\max\left\{\underline{m},\sum_{i\in R^{S}}Q_{i}\right\}$,   $q_{i}^{S}=Q_{i}$ for all $i\in R^{S}$,
and $q_{i}^{S}=0$ for $i\in S\setminus R^{S}$.

\end{description}
\end{theorem}

If a coalition of SMEs exports at optimality, all essential players therein, i.e., $S^{E}$, export their full capacity. If the combined capacity of the essential players is large enough to match or surpass the MQC, then no other players would be exporting. However, if the combined capacity of essential players falls short of MQC, it could be optimal for some complementary players to export as well.
Although these players do not have non-negative full capacity margins, their inclusion could be beneficial as they could reduce the under-supply penalty imposed to the essential players if they export on their own. 
Altogether, optimal export decisions of the complementary SMEs in a coalition depends on the capacities of the essential SMEs in that coalition. 
Theorem 1 reveals the intrinsic power of the essential players. If a player is essential, it always exports in a coalition which exports at optimality.  The optimal profit of a coalition is bounded by their aggregate full capacity export margins, as shown in the next corollary.
\begin{corollary}
\label{coro0}
    For every $S\subseteq N$ we have $\Pi(q^S,m^S)\leq \sum_{i\in S^{E}}\Delta_i $.
\end{corollary}
As shown by Theorem \ref{th3}, the exporting SMEs in a coalition only includes complementary players if the aggregate capacities of associated essential players is less than the MQC. In this case, however, one needs to find  the  optimal subset $D^S\subseteq S^{C}$  of complementary SMEs. 
We now discuss the optimisation problem for finding $D^S$. 
For $S\subseteq N$, and $D\subseteq S^{C}$ we define the function $G$ as
\begin{equation}
    \label{contcomp}
    G^{S}(D)= \sum_{i\in D}\Delta_i+\min\left\{\underline{m}-\sum_{i\in S^{E}}Q_i,\sum_{i\in D}Q_i\right\}r^u.
\end{equation}
The value of $G^S(D)$ is the contribution of complementary players in $D$ if they export along with the essential players $S^{E}$ in the export consortia of $S$. In coalitions that need complementary players, i.e., $S\subseteq N$  such that $\sum_{i\in S^{E}}Q_i<\underline{m}$, the problem of finding the optimal set of exporting  players is equivalent to maximizing the contribution of complementary players, as we show in the next result.
 
\begin{proposition}
\label{lemcompl}
    Given $S\subseteq N$, if $\sum_{i\in S^{E}}Q_i<\underline{m}$ and $R^{S}\neq \emptyset$, we have $R^{S}=S^{E}\cup D^{S}$ where $D^{S}\in \argmax_{D \subseteq S^{C}}G^{S}(D)$.    
\end{proposition}

For every coalition $S\subseteq N$, finding the optimal set of complementary SMEs $D^{S}\in \argmax_{D\subseteq S^{C}}G^S(D)$ readily obtains the optimal set of exporting players by adding the corresponding essential players, i.e., $R^{S}=S^{E}\cup D^{S}$. 
However, this optimization problem is NP-hard.

\begin{proposition}
\label{NPhard}
    The problem of finding  $D^{S}$  is NP-hard.
\end{proposition}

As the proof shows, the problem of finding $D^{S}$ (which is equivalent to the problem of finding $R^S$) is a variation of the $\{0,1\}$-knapsack problem which is known to be NP-hard \citep{pisinger1998knapsack}. 

As we have just demonstrated, identifying the optimal set of exporters for a specific coalition of SMEs is not a trivial matter. Furthermore, the reader may wonder whether this collaboration makes sense, i.e., if it is beneficial for all coalition members, and if there are benefit-sharing allocations that can be acceptable by all coalition members. To address this issue, the next section will define the cooperative games associated with such situations  and show that the cooperation among SMEs can always be made profitable for all the enterprises. 

\section{Cooperative Export Games}
 For each CE-situation $(N,Q,C,p,\underline{m})$, we can define an associated TU-game $%
(N,v)$, referred to as a cooperative export game (hereafter, a CE-game),
where the value of a coalition is the optimal profit of that coalition as calculated in (\ref{coalprof}) in the previous section. Thus, for $%
S\subseteq N$ the characteristic function is $v(S):=\Pi ^{S}(q^{S},m^{S})$. Note that the game can be explicitly defined as follows:
\begin{equation}
v(S)=%
\begin{cases}
\sum_{i\in R^{S}}\Delta_{i} -\left( \underline{m}-\sum_{i\in
R^{S}}Q_{i}\right) ^{+}r^{u} & \text{if }R^{S}\neq \emptyset, \\ 
0 & \text{if }R^{S}=\emptyset.%
\end{cases}%
\end{equation}
 where $R^{S}=\{i\in S:q_{i}^{S} > 0\}\subseteq S^{P}$ is the optimal set of exporting players  in $S$, and $v(\emptyset )=0.$ As we already mentioned in Section 6, $v(S) \geq 0.$  
Alternatively, for every $S\subseteq N$ such that $0<\sum_{i\in R^S}Q_i \leq \underline{m}$ we have $
      v(S)=\sum_{i\in S}\delta_i-\underline{m}r^u,$ and for every $S\subseteq N$ such that $\sum_{i\in R^S}Q_i > \underline{m}$ we have $
      v(S)=\sum_{i\in S}\Delta_i$.
      The following lemma highlights some properties of CE-situations that will be useful for the study of CE-games.

\begin{lemma}
\label{lemmma} Let $(N,Q,C,p,\underline{m})$ be a CE-situation.  Let $S^{0}\subseteq N^{E}$ be an arbitrary subset of
essential players and $S=S^{0}\cup N^{C}$. If $R^{S},R^{N}\neq
\emptyset $. It holds that

\begin{description}
\item[(i)] $\sum_{i\in
D^{S}}\Delta _{i}\leq \sum_{i\in D^{N}}\Delta _{i}.$


\item[(ii)] 
$\sum_{i\in D^{S}}\Delta _{i} -\left( \underline{m}-\sum_{i\in R^{S}}Q_{i}\right)
^{+}r^{u}\leq 
\sum_{i\in
D^{N}}\Delta _{i} -
\left( \underline{m}-\sum_{i\in R^{N}}Q_{i}\right) ^{+}r^{u}
.
$

\end{description}
\end{lemma}

Recall that $D^S$ is the optimal set of complementary exporting SMEs in $S$. 
Property (i) asserts that the total full capacity margin of optimal complementary players in sub-coalitions, which include a subset of essential players, is never larger than that in the grand coalition. Considering that full capacity margins are negative for complementary players, in absolute terms, the grand coalition employs a set of complementary players with a smaller or equal full capacity margin.
Property (ii) is a technical observation which helps in proving the subsequent results.  The next proposition elucidates the main properties of CE-games, including superadditivity. 
 
\begin{proposition}
\label{Lemma game} Let $(N,Q,C,p,\underline{m})$ be a CE-situation, and let $%
(N,v)$ be the corresponding CE-game. The following statements hold:
\begin{description}
\item[(i)] $v(T)=0$ for all $T\subseteq N\setminus N^{E}.$

\item[(ii)] $v(S\setminus T)=v(S)$ for all $T\subseteq S\setminus
S^{P}.$

\item[(iii)] \label{supad}$v(S\cup T)\geq v(S)+v(T)$ for all disjoint sets $%
S,T\subseteq N.$

\item[(iv)] $v(S)\leq v(T)$ for all $S\subseteq T\subseteq N.$
\end{description}
\end{proposition}


Property (i) indicates that in coalitions where there are no
essential players, there is no export. Meanwhile, property (ii) asserts
that players who do not export under any circumstances contribute no benefit
to the coalition. Property (iii) states that CE-games are superadditive.
Consequently, the formation of the grand coalition is justified. On the
other hand, property (iv) shows that the game is monotone increasing. 

Before exploring the allocation of profits from this cooperation among the
players, we illustrate a three-player CE-situation and its associated game with an example.

\begin{example}
\label{ex1}
    Consider a scenario with three players $N=\{1,2,3\}$. We have $Q_1=14$ , $Q_2=33$, $Q_3=21$, $c_1=15$, $c_2=7$, $c_3=23$. Let $p=21$, $\underline{m}=61$, and $r^u=10$. The coalitions' optimal profits and strategies are shown in Table \ref{tab:ex1t}.

\begin{table}[t]
\begin{center}
\begin{tabular}{c| c c}
\hline
      $S$        &         $R^{S} | (S\setminus R^{S})$ & $v(S)$ \\
     \hline
    $  \{1\}$            &       $\emptyset|\{1\}$ & 0\\    
    $  \{2\}$             &       $\{2\}|\emptyset$ & 406\\
    $  \{3\}$             &        $\{3\}|\emptyset$ & 18\\
    $  \{1,2\}$           &       $\{1,2\}|\emptyset$ & 825\\
    $  \{1,3\}$           &       $\{1,3\}|\emptyset$ & 437\\
    $  \{2,3\}$           &       $\{2,3\}|\emptyset$ & 1034\\
    $  \{1,2,3\}$        &       $\{1,2,3 \}|\emptyset$ & 1383\\
    \hline
\end{tabular}  
\end{center}
\caption{The coalitions' optimal strategies and profits in Example \ref{ex1}}
\label{tab:ex1t}
\end{table}
   All players are essential, but player 1 cannot export profitably on his own unlike the other two players who could do so individually. Any coalition that is formed of two or more players will export  profitably.  Note that this game is not convex: $ 349=v(\{1,2,3 \})-v(\{2,3 \}) < v(\{1,2 \})-v(\{2 \})=419$. We can find stable allocations in the core for this game. For instance, distributing the benefits of the grand coalition proportionally to $\Delta_{i}$ among all essential players, that is $(279,686,418)$, gives an allocation in the core. 
\end{example}
 
The CE-games are not generally convex as seen in Example \ref{ex1}, thus the Shapley value \citep{shapley:book1952} is not necessarily in the core (in addition to being computationally challenging). We now introduce the NEA-allocation rule which is always in the core and also computationally easy to calculate. 

Let $(N,Q,C,p,\underline{m})$ be a CE-situation, and let $(N,v)$ be the
corresponding CE-game, we define the allocation $\phi ( N,v )=(\phi
_{i}(N,v))_{i\in N}$   as follows:
\begin{equation}
\label{alloc}
\phi _{i}(N,v):=%
\begin{dcases}
\frac{\Delta _{i}}{\sum_{j\in N^{E}}\Delta _{j}}v(N)& \text{if }i\in N^{E}
\\ 
0 & \text{otherwise}%
\end{dcases}%
\end{equation}
The above allocation rule divides the gains of the grand coalition among the essential players proportional to their individual full capacity margins. All other players will receive zero benefit from cooperative exporting. This includes the the complementary SMEs who do export along with the essential players, i.e., $D^{N}$. So, complementary SMEs' participation in the export consortia has to be without any expectations for receiving a positive gain. Due to this feature, we will call this allocation Non-essential Exporter Altruistic rule (henceforth NEA rule).

The next theorem shows that the NEA rule is coalitionally stable in the sense of the core.  

\begin{theorem}
\label{th4} Let $(N,Q,C,p,\underline{m})$ be a CE-situation, and let $(N,v)$
be the corresponding CE-game. Then, NEA rule, as defined in equation (\ref{alloc}), belongs to the core of the game.
\end{theorem}
Following  the above theorem, CE-games are balanced. It is easy to see that every subgame of CE-games is also a CE-game. Thus, CE-games are totally balanced.

As illustrated by Example \ref{ex1}, in situations where complementary players are absent, the NEA rule emerges as a fair method for distributing the benefits generated by the grand coalition. Nevertheless, in the subsequent scenario involving the same number of SMEs, it becomes evident that not all exporting players see the benefits of their cooperation.

    
\begin{example}
\label{exM}
   Consider a scenario with three players $N=\{1,2,3\}$. We have $Q_1=5$ , $Q_2=5$, $Q_3=6$, $c_1=20$, $c_2=20$, $c_3=35$. Let $p=5.5$, $\underline{m}=11$, and $r^u=6$. The coalitions' optimal profits and strategies are shown in Table \ref{tab:exM}.

\begin{table}[t]
\begin{center}
\begin{tabular}{c| c c}
\hline
      $S$        &        $R^{S} | (S\setminus R^{S})$ & $v(S)$ \\
     \hline
    $  \{1\}$           &       $\emptyset|\{1\}$ & 0\\    
    $  \{2\}$            &       $\emptyset|\{2\}$ & 0\\
    $  \{3\}$             &        $\emptyset|\{3\}$ & 0\\
    $  \{1,2\}$           &       $\{1,2\}|\emptyset$ & 9\\
    $  \{1,3\}$           &       $\{1,3\}|\emptyset$ & 5.5\\
    $  \{2,3\}$           &       $\{2,3\}|\emptyset$ & 5.5\\
    $  \{1,2,3\}$         &       $\{1,2,3 \}|\emptyset$ & 13\\
    \hline
\end{tabular}  
\end{center}
\caption{The coalitions' optimal strategies and profits in Example \ref{exM}}
\label{tab:exM}
\end{table}
In this example, players 1 and 2 are essential while player 3 is complementary. Although player 3 exports  with either 1 or 2, as well as in the grand coalition, the NEA rule distributes the profit in the following way: $\phi (N,v)=(6.5,6.5,0)$, allocating no profit to player 3. Nevertheless, there still exists alternative allocations in the core that could give  as much as 4 units of profit to player 3 (for example, consider the allocation $(4.5,4.5,4)$.
\end{example}
To address the incentive issues of the complementary exporting players, a  possible alternative to the NEA rule would be to consider distributing the benefits of the grand coalition proportionally to $\delta _{i}$ for all exporters (essential and complementary). Consider then the following proportional allocation:   
$$ 
\omega _{i}(N,v):=%
\begin{dcases}
\frac{\delta _{i}}{\sum_{j\in R^{N}}\delta _{j}}v(N)& \text{if }i\in R^N
\\ 
0 & \text{otherwise}%
\end{dcases}%
$$
Here, all exporting players receive  an allocation proportional to their $\delta_i$,  and therefore we will call it $\delta$-proportional rule. Since all exporting players are compensated based on the same criteria, there is a natural `fairness' intuition associated with this solution. Unfortunately, this allocation is not necessarily in the core. A counterexample can be observed in Example \ref{exM} where we get $\omega (N,v)=(4.47,4.47,4.06)$. With this allocation we have $v(\{1,2\})=8.94<9$ which violates stability condition for this coalition.  
Considering the nucleolus allocation for this example, we get $\eta(N,v)=(5.5,5.5,2)$. It removes one unit from each essential player and gives it to the complementary exporting player. The nucleolus also obtains a stable allocation in the sense of the core for these games, which are balanced. However, as the number of players increases, computing the nucleolus becomes computationally challenging. \citet{perea2019heuristic} proposes a heuristic approach for calculating the nucleolus, which is based on sampling the coalitions space.  

Although the NEA rule is in the core and it is computationally straightforward, it does not give any extra value to the complementary players that are part of the optimal exporters in $N$. Thus, one can argue that the NEA rule does not provide sufficient incentives for complementary exporting players to participate in the coalition even when their help is needed to achieve the economy of scale required for export. We address this issue next by refining the NEA rule by means of compensatory policies for complementary exporting players. 
In the following section, we introduce two alternative allocations, easily computable, aimed at compensating complementary exporting players for their contributions to the collective export benefit.

\section{Compensation policies for complementary exporting players}

The NEA rule is altruistic towards complementary exporting players since it
don't give away any margins to these players. Thus, one remains dubious
about the willingness of these players to join export consortia. In what
follows, we give two modifications to this allocation rule to remedy this
drawback. These allocations share a portion of export margins among
complementary exporting players by taxing essential players. The first allocation rule
does this by charging all essential players a fixed amount. The second
allocation rule, on the other hand, charges an amount proportional to the
gain received by each essential player. 

\subsection{Egalitarian essential rate}
The first approach to modify the NEA allocation rule in (\ref{alloc}) involves charging a fixed rate from the essential SMEs in the coalition and distribute the collected amount among the complementary exporting players. 
Let $(\alpha _{i})_{i\in D^{N} }$ be a system of  non-negative  weights for
complementary exporting players in the grand coalition.   That is, $\alpha _{j} \geq 0$ for all $j \in D^{N}$ with at least a $k \in D^{N}$ such that $\alpha _{k} > 0$.  Given  $\rho \geq 0$, the \textit{egalitarian essential rate}  allocation rule $\psi ^{\rho }(N,v)=(\psi _{i}^{\rho
}(N,v))_{i\in N}$ is defined as follows: 
\begin{equation}
\label{alloc2}
\psi _{i}^{\rho }(N,v)=%
\begin{dcases}
\phi _{i}(N,v)-\rho  & \text{if }i\in N^{E} \\ 
\frac{\alpha _{i}}{\sum_{j\in D^{N}}\alpha _{j}}|N^{E}| \rho
& \text{if }i\in D^{N}  \\ 
0 & \text{if }i\in N\setminus R^{N}%
\end{dcases}%
\end{equation}
 The egalitarian essential rate  rule charges a fix amount $\rho \geq 0$ from every essential player
after allocating gains among players according the NEA allocation rule $%
\phi (N,v)$.

Our next objective is to find out what values of $\rho $ (tax rate) are
affordable for the essential players so that the associated allocation $\psi ^{\rho
}(N,v)$ is within the core. To do so, we will use the excess
function for coalitions  under NEA rule $\phi (N,v)$. That is, for a coalition $S\subset N$, let $%
e(S,\phi (N,v))=\sum_{i\in S}\phi _{i}(N,v)-v(S)$. Note that  since $\phi (N,v)\in
Core(N,v)$ we have $e(S,\phi (N,v))\geq 0$ for all $S\subseteq N$. We are particularly interested in the coalition $\hat{S}\subset N$ that attains the minimum 
excess per number of essential players it
contains, that is: 
\begin{equation}
\hat{S}\in \argmin_{S\subsetneq N:S^{E}\neq \emptyset }  \frac{%
e(S,\phi (N,v))}{|S^{E}|}.
\end{equation}%
As per formulation above, finding the coalitions with minimum excess per number of essential requires searching among all sub-coalitions which, as the number of players grow, can be computationally challenging. However, as our next result shows, such sub-coalitions can only be among certain groups of coalitions. Consider $M_{i}(N,v)=v(N)-v(N\setminus \{i\})$ as the
marginal contribution of player $i$.
\begin{lemma}
\label{lemexc} Let $(N,Q,C,p,\underline{m})$ be a CE-situation, and let $%
(N,v)$ be the corresponding CE-game. Given the NEA rule $\phi (N,v)$, we have
either $\hat{S}=N\setminus \{i^{\ast }\}$ for $i^{\ast }\in N$ such that $%
M_{i^{\ast }}(N,v)-M_{i}(N,v)\leq \phi _{i^{\ast }}(N,v)-\phi _{i}(N,v)$ for all $i \in N$, or $\hat{S}%
=(N\setminus N^{E})\cup \{i^{\ast }\}$ for $i^{\ast }\in \argmin_{i\in
N^{E}}\Delta _{i}$.
\end{lemma}
The first possibility for the sub-coalition with minimum excess per number of essential players is the coalition which has only one essential player less than the grand coalition. The absent player,  $i^{\ast }$, in this case is the essential player who receives the closest allocation to his/her maximum possible allocatable gain under the original NEA rule. Note that in this case we have $|\hat{S}|\in \{1,n-1\}$. The second possibility for the sub-coalition with minimum excess per number of essential players is the coalition which includes all non-essential players and only the essential player with the minimum full capacity export margin. Figuring out $\hat{S}$ allows us to calculate the maximum fix rate that can be deducted from the essential players to be redistributed among the complementary exporting players. The following result provides a sufficient condition for the egalitarian
essential rate rule to belong to the core of the game.

\begin{theorem}
\label{th5} Let $(N,Q,C,p,\underline{m})$ be a CE-situation, and let $(N,v)$
be the corresponding CE-game. If $D^{N}\neq \emptyset $, for
every $0\leq \rho \leq \frac{e(\hat{S},\phi (N,v))}{|\hat{S}^{E}|}$ the allocation
rule $\psi ^{\rho }(N,v)$ is in the core of the associated game.
\end{theorem}
In light of the previous result, we have an interval to select an appropriate $\rho$ within which  so that the egalitarian essential rate allocation is stable in the sense of the core. However, we can set further criteria for choosing the appropriate $\rho$, which more `fairly' compensates complementary exporting players. Therefore, we  establish the proportional allocation in (\ref{alloc2}) as a reference for compensation for complementary exporting players, which, despite not belonging to the core, distributes profit proportionally to the under-supply adjusted (full capacity) margin of exporting players, and in that sense treats all players in a same way.
With this objective in mind, we let $\alpha_i=\delta_i$ for all $i\in D^N$, and define 
\begin{equation}
    \rho^E:= \min \left\lbrace \frac{e(\hat{S},\phi (N,v))}{|\hat{S}^{E}|}, \frac{ v(N)   \sum_{j\in D^{N}}\delta _{j}}{|N^{E}|  \sum_{j\in R^{N}}\delta _{j}}  \right\rbrace.
\end{equation}
One can observe that in this way, if the the second expression is less than or equal to the first expressions, then the egalitarian essential allocation generates the same allocations for complementary exporting players as that in  (\ref{alloc2}) while ensuring that the resultant allocation is in the core. 

\subsection{Proportional essential rate}
The second approach to modify the NEA allocation rule in (\ref{alloc}) is to tax the essential SMEs in the coalition (with a fix percentage rate) and distribute the collected amount among the complementary exporting players. 
Let $(\alpha _{i})_{i\in D^{N}}$ be a system of non-negative  weights for
complementary exporting players in the grand coalition. Given  $\rho \geq 0$, the \textit{proportional essential rate} allocation rule $\varphi ^{\rho }(N,v)=(\varphi
_{i}^{\rho }(N,v))_{i\in N}$ is defined as follows: 
\begin{equation*}
\varphi _{i}^{\rho }(N,v)=%
\begin{dcases}
(1-\rho )\phi _{i}(N,v) & \text{if }i\in N^{E} \\ 
\frac{\alpha _{i}}{\sum_{j\in D^{N}}\alpha _{j}}\rho v(N) & 
\text{if }i\in D^{N} \\ 
0 & \text{if }i\in N\setminus R^{N}%
\end{dcases}%
\end{equation*}
The proportional essential rate allocation rule charges an amount $\phi _{i}(N,v)\rho \geq 0$ from every essential player, after allocating gains among players according the NEA rule 
$\phi (N,v)$, and then redistributes the aggregated amount among the complementary exporting players according to the weights $(\alpha _{i})_{i\in D^{N}}$.

Consider the ratio of the excess value under the NEA rule assigned to a coalition which contains at least an essential player, and the aggregated allocation of that coalition under the NEA rule. 
Among these values of all feasible sub-coalition, we highlight the sub-coalition with the minimum ratio:  
\begin{equation}
    \check{S}=\argmin_{S\subsetneq N:S^{E}\neq \emptyset }\frac{e(S,\phi (N,v))}{\sum_{i\in S }\phi _{i}(N,v)}.
\end{equation}
The following result shows that the coalition where the ratio is
minimised, $\hat{S},$ can be found amongst a restricted set of possible sub-coalitions. 
\begin{lemma}
\label{lem4}  Let $(N,Q,C,p,\underline{m})$ be a CE-situation, and let $(N,v)
$ be the corresponding CE-game. Given the NEA rule $\phi (N,v)$, then $\check{S%
}=N\setminus \{i^{\ast }\}$ where $i^{\ast }\in \argmin_{i\in N^{E}}\frac{e(N\setminus \{i\},\phi (N,v))}{\sum_{j\in N\setminus \{i\} }\phi _{j}(N,v)}.$
\end{lemma}
In fact, $\check{S}$ has exactly one essential player less than the grand coalition $N$. Thus, finding the latter sub-coalition is computationally straightforward. 
Our next result indicates a range for $\rho$ that results in  the proportional essential rate allocation rule to belong to the core of the game.
\begin{theorem}
\label{th6} Let $(N,Q,C,p,\underline{m})$ be a CE-situation, and let $(N,v)$
be the corresponding CE-game. If $D^{N}\neq \emptyset $, for
every $0\leq \rho \leq  \frac{e(\check{S},\phi (N,v))}{\sum_{i\in \check{S} }\phi _{i}(N,v)}$ the allocation rule $\varphi ^{\rho }(N,v)$ is in the core of the associated game.
\end{theorem}
A similar argument to the previous subsection allows us to define a particular  value for $\rho$ such that the proportional essential rate allocation rule would belong to the core and approximates the compensation provided by  the $\delta$-proportional rule  for complementary exporting players. For all $i\in D^N$ let $\alpha_i=\delta_i$, and define
\begin{equation}
  \rho^P:= \min \left\lbrace \frac{e(\check{S},\phi (N,v))}{\sum_{i\in \check{S} }\phi _{i}(N,v)}, \frac{  \sum_{j\in D^{N}}\delta _{j}}{\sum_{j\in R^{N}}\delta _{j}}  \right\rbrace.  
\end{equation} 
 The reader may notice that $0\leq \rho^P < 1$. One can check that when the upper bound of $\rho$ is large enough, $\rho^P$ obtains allocations for the complementary exporting players that are equal to what allocation (\ref{alloc2}) obtains for these players. 
We comparing our modified  allocation rules through numerical experiments in the next section.

\section{Numerical Experiments}
In this section, we compare the two modified solutions proposed in the previous section to examine which provides a greater compensation for complementary exporting players. We start by considering a CE-situation with four players.

\begin{example}
\label{ex3}
 Consider a scenario with four players $N=\{1,2,3,4\}$. We have $Q_1=10$ , $Q_2=15$, $Q_3=15$, $Q_4=30$, $c_1=1$, $c_2=c_3=25$, $c_4=200$. Let $p=6$, $\underline{m}=50$, $r^u=5$ and $\alpha _{i}=\delta_i$ for all $i \in D^{N}$. The coalitions' optimal profits and strategies are shown in Table \ref{tab:ex2t}.
\begin{table}[t]
\begin{center}
\begin{tabular}{c| c c}
\hline
      $S$        &         $R^{S} | (S\setminus R^{S})$ & $v(S)$ \\
     \hline
    $  \{1\}$          &       $\emptyset|\{1\}$ & 0\\    
    $  \{2\}$           &       $\emptyset|\{2\}$ & 0\\
    $  \{3\}$            &        $\{3\}|\emptyset$ & 0\\
    $  \{4\}$             &        $\emptyset|\{4\}$ & 0\\
    $  \{1,2\}$           &       $\{1,2\}|\emptyset$ & 0\\
    $  \{1,3\}$           &       $\{1,3\}|\emptyset$ & 0\\
    $  \{1,4\}$           &       $\emptyset|\{1,4\}$ & 0\\
    $  \{2,3\}$           &       $\{2,3\}|\emptyset$ & 30\\
    $  \{2,4\}$           &       $\emptyset|\{2,4\}$ & 20\\
    $  \{3,4\}$           &       $\{3,4\}|\emptyset$ & 20\\
    $  \{1,2,3\}$         &       $\{1,2,3 \}|\emptyset$ & 139\\
    $  \{1,2,4\}$        &       $\{1,2,4 \}|\emptyset$ & 104\\
    $  \{1,3,4\}$         &       $\{1,3,4 \}|\emptyset$ & 104\\
    $  \{2,3,4\}$         &       $\{2,3,4 \}|\emptyset$ & 110\\
    $  \{1,2,3,4\}$      &       $\{1,2,3,4 \}|\emptyset$ & 169\\
    \hline
\end{tabular}  
\end{center}
\caption{The coalitions' optimal strategies and profits in Example \ref{ex3}}
\label{tab:ex2t}
\end{table}

In this example, players 1 is extremely efficient, with a very low cost. But he does not have sufficient quantity. Players 2 and 3 are symmetric with mid levels of efficiency and quantity. These three players are essential. Player 4 has the largest quantity but he is extremely inefficient, i.e., with a high cost. Thus player 4 is complementary. 
Table \ref{tab:ex3t} exhibits the comparison among the rules studied in this research as well as the nucleolus.

\begin{table}
\begin{center}
\begin{tabular}{c|c|c|c|c}
$\phi(N,v)$ &  $\omega(N,v)$  & $\psi ^{\rho^E }(N,v)$ & $\varphi ^{\rho^P}(N,v)$ & $\eta(N,v)$ \\ \hline
$\left( 
\begin{array}{c}
52.75 \\ 
58.125 \\ 
58.125 \\ 
0%
\end{array}%
\right) $ &
 $\left( 
\begin{array}{c}
35.50 \\ 
45.60 \\ 
45.60 \\ 
42.20%
\end{array}%
\right) $
&

$\left( 
\begin{array}{c}
49.65 \\ 
55 \\ 
55 \\ 
9.35%
\end{array}%
\right) $ & $\left( 
\begin{array}{c}
49.95 \\ 
55 \\ 
55 \\ 
9.05%
\end{array}%
\right) $ & $\left( 
\begin{array}{c}
46.5 \\ 
52.5 \\ 
52.5 \\ 
17.5%
\end{array}%
\right) $
\end{tabular}%
\caption{Comparison of the different solutions proposed in Example \ref{ex3}}
\label{tab:ex3t}
\end{center}
\end{table}

\end{example}

Player 4 has an extremely high cost compare to the other three players so $\Delta_4<0$ and as such this is a complementary player. In the end, the NEA allocation rule gives him nothing, although the $\delta$-proportional rule gives this player as high as 42.
Our allocation rules recommend a maximum of   $\rho^E=3.12$ egalitarian payment from the three essential players 1,2,3, or $\rho^P=0.05 $ of their NEA allocation back as tax ratio to compensate player 4. This gives the player 4 a maximum of 9.35 units of gained profit.

 To exhibit the burden of player 4 to the rest of the players, consider an alternative situation where an additional efficient player 5, with same quantity as player 4, $Q_5=Q_4$, but $c_5=5$. Then the optimal set of exporters does not include player 4 and the essential players have increased their gains (under NEA allocation rule) to  $\phi^{\prime}_1=59$, $\phi^{\prime}_2=65$, and $\phi^{\prime}_3=65$, that is around $15\%$ increase on average. 
 

\begin{example}
\label{ex4}
    Consider a scenario with four players $N=\{1,2,3,4\}$. We have $Q_1=8$ , $Q_2=9$, $Q_3=54$, $Q_4=37$, $c_1=c_3=10$, $c_2=28$, $c_4=15$. Let $p=3$, $\underline{m}=103$, $r^u=12$ and $\alpha _{i}=\delta_i$ for all $i \in D^{N}$. In this example, players 1, 3 and 4 are essential and player 2 is complementary. The non-zero values of the objective function are: $v(\{3,4\})=104$, $v(\{1,3,4\})=214$, $v(\{2,3,4\})=211$ and $v(\{1,2,3,4\})=261$. Moreover, $\rho^E=5.9788$ and $\rho^P=0.0687 $. The comparison of the different solutions is presented in Table \ref{tab:ex4t}.

\begin{table}
\begin{center}
\begin{tabular}{c|c|c|c|c}
$\phi(N,v)$ & $\omega(N,v)$  & $\psi ^{\rho^E }(N,v)$ & $\varphi ^{\rho^P}(N,v)$ & $\eta(N,v)$ \\ \hline
& & & \\ 
$\left( 
\begin{array}{c}
13.95 \\ 
0 \\ 
151.42 \\ 
95.63%
\end{array}%
\right) $ &
 $\left( 
\begin{array}{c}
18.44 \\ 
17.93 \\ 
134.10 \\ 
90.53%
\end{array}%
\right) $& 
$\left( 
\begin{array}{c}
7.97 \\ 
17.93 \\ 
145.44 \\ 
89.66%
\end{array}%
\right) $ & $\left( 
\begin{array}{c}
12.99 \\ 
17.93 \\ 
141.01 \\ 
89.07%
\end{array}%
\right) $ & $\left( 
\begin{array}{c}
25 \\ 
23.5 \\ 
106.25 \\ 
106.25%
\end{array}%
\right) $
\end{tabular}%
\caption{Comparison of the different solutions proposed in Example \ref{ex4}}
\label{tab:ex4t}
\end{center}
\end{table}

\end{example}

\begin{example}
\label{ex5}
    Consider a scenario with four players $N=\{1,2,3,4\}$. We have $Q_1=43$ , $Q_2=12$, $Q_3=4$, $Q_4=3$, $c_1=9$, $c_2=17$, $c_3=21$, $c_4=6$. Let $p=5$, $\underline{m}=68.5$, $r^u=18$ and $\alpha _{i}=\delta_i$ for all $i \in D^{N}$. In this example players 1, 2 and 4 are essential and player 3 is complementary. The non-zero values of the objective function are: $v(\{1,2\})=6$, $v(\{1,2,3\})=77$, $v(\{1,2,4\})=69$ and $v(\{1,2,3,4\})=140$. Moreover, $\rho^E=2.4132$ and $\rho^P=0.0517 $. The comparison of the different solutions is presented in Table \ref{tab:ex5t}.

\begin{table}
\begin{center}
\begin{tabular}{c|c|c|c|c}
$\phi(N,v)$ & $\omega(N,v)$  & $\psi ^{\rho^E }(N,v)$ & $\varphi ^{\rho^P}(N,v)$ & $\eta(N,v)$ \\ \hline
& & & \\ 
$\left( 
\begin{array}{c}
111.78 \\ 
23.33 \\ 
0 \\ 
4.89%
\end{array}%
\right) $ &
 $\left( 
\begin{array}{c}
99.93 \\ 
26.41 \\ 
7.24 \\ 
6.42%
\end{array}%
\right) $ & 
$\left( 
\begin{array}{c}
109.37 \\ 
20.92 \\ 
7.24 \\ 
2.47%
\end{array}%
\right) $ & $\left( 
\begin{array}{c}
106 \\ 
22.13 \\ 
7.24 \\ 
4.63%
\end{array}%
\right) $ & $\left( 
\begin{array}{c}
36.5 \\ 
36.5 \\ 
35.5 \\ 
31.5%
\end{array}%
\right) $
\end{tabular}%
\caption{Comparison of the different solutions proposed in Example \ref{ex5}}
\label{tab:ex5t}
\end{center}
\end{table}

\end{example}

 Note that in example 3, the nucleolus is close to the proposed alternative solutions. However, in examples 4 and 5, the nucleolus exhibits a very different behavior, as it penalizes an essential player. Specifically, in example 5 the nucleolus punishes player 1 with more than $50\%$ of their profit while compensating not only the complementary player (player 3) but also the rest of the essential players (players 2 and 4). In example 4, it harms the essential player 3 with more than $25\%$ of their profit. We would also like to note that the $\delta$-proportional rule is a core allocation for examples 4 and 5, but not for examples 2 and 3. 


 \section{Conclusions}

  International trade presents significant opportunities for agri-food communities to generate profits. However, for small and medium-sized enterprises (SMEs), accessing international markets can be financially prohibitive, particularly under trade policies that impose minimum quantity commitments (MQCs) on export volumes, such as licensing tariff rate quota (TRQ) mechanisms.

In our study, we aim to address this challenging context with the goal of developing optimal strategies to facilitate international market exportation for Small and Medium-sized Enterprises (SMEs). The central idea is to design collaborative approaches, such as the formation of export consortia, that enable SMEs to join forces to overcome inherent barriers in global competition. In this regard, the implementation of strategies focused on export commitment and the optimization of the agri-food supply chain will be crucial. By fostering collaboration and synergy between SMEs, we aim to contribute to a more robust and competitive commercial ecosystem where both types of businesses can mutually thrive in the international market. Adaptability in the face of global uncertainty will be a key element in our proposals, ensuring the long-term viability of these strategies in a dynamic business environment.

This paper demonstrates how cooperative exporting among agri-food SMEs can effectively overcome the challenges posed by MQCs, thereby expanding market access to a wider array of SMEs. By formulating a class of cooperative games tailored to these scenarios, we identify a gain-sharing mechanism that ensures allocations within their respective cores, facilitating the formation of stable grand coalitions among cooperative exporting SMEs.

Our proposed allocation NEA-rule distributes the export surplus exclusively among the ``essential'' SME exporters—those that demonstrate sufficient cost efficiency. Consequently, less cost-efficient ``complementary'' SMEs, whose capacities are essential for meeting MQCs, do not directly benefit from collaborative exporting and must participate altruistically.

To address this issue, we suggest two modifications to our original allocation rule aimed at sharing a portion of the export surplus with complementary SMEs through taxing the essential SMEs. These modifications include an egalitarian approach, which we called Egalitarian essential rate, and a revenue-based rate approach, called Proportional essential rate. Through numerical examples, we compare the performance of these allocation mechanisms and discuss their practical implications for cooperative exporting strategies in agri-food communities.

Future research in this domain could explore several avenues to further enhance our understanding and implementation of cooperative exporting strategies among agri-food SMEs. Firstly, investigating attainable allocations through the lens of a potential-based mechanism (PMAS) could offer insights into more dynamic and adaptive gain-sharing approaches, taking into account the evolving needs and capacities of participating SMEs. Secondly, delving into detailed characterizations of the proposed solutions, including their stability, fairness, and scalability under various market conditions and policy frameworks, would provide a deeper understanding of their practical implications and limitations. Finally, applying these cooperative exporting models to real-world data from agri-food exporting companies would offer empirical validation and refinement, enabling the development of more tailored and effective strategies to promote inclusive and sustainable international trade within the agricultural sector. 

\section*{Acknowledgements}
This work is part of the R+D+I project grants PID2022-137211NB-100,
that were funded by MCIN/AEI/10.13039/501100011033/ and by
‘‘ERDF A way of making Europe’’/EU. This research was also funded
by project PROMETEO/2021/063 from the Conselleria d'Innovació, Universitats, Ciència i Societat Digital, Generalitat Valenciana.
 
\bibliography{bibliography}
\bibliographystyle{apalike}

   \newpage
\section*{Appendix - Proofs}
\begin{proof}[Proof of Lemma \ref{th1}]
    It is straightforward to verify above since without any restriction on $m_i$, and non-negative deviation penalties, it is always optimal to match export quantity with the commitment level, that is, $q^*_i=m^*_i$. This simplifies the profit function to $\Delta_i \mathbf{1}_{q_i>0}$ and subsequently we get the two possibilities as stated above.
\end{proof}

\begin{proof}[Proof of Lemma \ref{th2}]
If $Q_i\geq \underline{m}$, the MQC would not affect the optimal solution and, similar to the case in the previous section, we get $q^*_i=m^*_i=Q_i$ if $\Delta_i\geq 0$ and $q^*_i=m^*_i=0$ otherwise. 
Thus, for the rest of the proof we assume $Q_i< \underline{m}$.

Since $Q_i< \underline{m}$, upon exporting there will  be under supply penalty. Thus the best choice of commitment if $q_i>0$ is $m^*_i=\underline{m}$. The profit in this case is
$$
\Pi(q_i,\underline{m})=q_i(p+r^u)-c_i-\underline{m}r^u
$$
The first part is increasing on $q_i$ thus maximum profit is 
$Q_i(p+r^U)-c_i-\underline{m}r^u$. If this value is positive, that is, $\delta_i\geq \underline{m}r^u$, then we have $q^*_i=Q_i$. Otherwise, we have $q^*_i=m^*_i=0$.
\end{proof}

\begin{proof}[Proof of Theorem \protect\ref{th3}]
Let $R^{S}\subseteq S$ be the subset of players in $S$ such as $q^S_i>0$ for all $i \in R^{S}$. We
assume that cooperating via the export coalition is beneficial, that is, $%
R^{S}\neq \emptyset$. By Lemma \ref{th2}, we have $q^S_i=Q_i$ for all $i\in
R^{S}$. We consider the cases in the statement of the theorem in sequence:\newline

\noindent (A) ``If $\sum_{i\in S^{E}}Q_{i}\geq  \underline{m}$ then $m^{S}=\sum_{i\in S%
^{E}}Q_{i}$,   $q_{i}^{S}=Q_{i}$ for all $i\in S^{E}$,
and $q_{i}^{S}=0$ for $i\in S\setminus S^{E}$."

We first show that $S^{E}\subseteq R^{S}$. Since $m^{S}\geq \underline{m}
$ the profit function at optimality is 
\begin{equation*}
\Pi ^{S}(q^{S},m^{S})=\sum_{i\in R^{S}}\Delta_i 
\end{equation*}

Suppose the contrary, that is, there exist no optimal solution $(q^S,m^S)$
where $q_i^S=Q_i$ for all $i\in S^{E}$. In this case, let $j\in S$ be a player
such that $j\in S^{E}$ but $j\notin R^{S}$. Consider an alternative solution
with $q_{j} =Q_{j}$ , $q_i=q^S_i$ for all $i\in S\setminus \{j\}$, and $%
m=m^S+Q_j$. We have $\Pi^S(q,m)- \Pi^S(q^S,m^S)=\Delta_j\geq 0$ which is a
contradiction. Thus $S^{E}\subseteq R^{S}$.

Next, we show that $S^{E}=R^{S}$. Suppose $S^{E}\subsetneq R^{S}$, and let $j\in
R^{S}\setminus S^{E}$. Consider an alternative solution with $q_{j}=0$, $%
q_i=q^S_i$ for all $i\in S\setminus \{j\}$, and $m=m^S-Q_j$. Note that $m> 
\underline{m}$. We have $\Pi^S(q,m)- \Pi^S(q^S,m^S)=\Delta_j<0$ which is a
contradiction. Thus $S^{E}= R^{S}$.\newline

\noindent (B) ``If $\sum_{i\in S^{E}}Q_i< \underline{m}$ then 
for some $R^{S}\subseteq S$ such that $S^{E}\subseteq R^{S}\subseteq S^{P}$, we have $m^{S}=\max\left\{\underline{m},\sum_{i\in R^{S}}Q_{i}\right\}$,   $q_{i}^{S}=Q_{i}$ for all $i\in R^{S}$,
and $q_{i}^{S}=0$ for $i\in S\setminus R^{S}$."

We first show that $S^{E}\subseteq R^{S}$. The optimal profit in this
case is 
\begin{equation*}
\Pi ^{S}(q^{S},m^{S})=\sum_{i\in R^{S}}\left[ Q_{i}(p+r^{u})-c_{i}\right]
-m^{S}r^{u}.
\end{equation*}%
Suppose the contrary, that is, there exist no optimal solution $(q^{S},m^{S})
$ where $q_{i}^{S}=Q_{i}$ for all $i\in R^{S}$. In this case, let $j\in S$
be a player such that $j\in S^{E}$ but $j\notin R^{S}$. There are different
possibilities:

\begin{enumerate}
\item $Q_j<m^S-\sum_{R^{S}}Q_i$: Consider an alternative solution with $q_{j}
=Q_{j}$ , $q_i=q^S_i$ for all $i\in S\setminus \{j\}$, and $m=m^S=\underline{m%
}$. We have $\Pi^S(q,m)- \Pi^S(q^S,m^S)=\delta_j\geq0$ which is a
contradiction.

\item $Q_j\geq m^S-\sum_{R^{S}}Q_i$: Consider an alternative solution with $%
q_{j} =Q_{j}$, $q_i=q^S_i$ for all $i\in S\setminus \{j\}$, and $%
m=\sum_{i\in R^{S}\cup \{j\}}Q_i$. We have $\Pi^S(q,m)-
\Pi^S(q^S,m^S)=\Delta_j+\left(\underline{m}-\sum_{i\in R^{S}}Q_i\right)r^u\geq
0$ which is a contradiction.
\end{enumerate}

Thus $S^{E}\subseteq R^{S}$.

Next, we show that there is no $j\in R^{S}$ such that $j\in S\setminus S^{P}$%
. Suppose the contrary, that is, let $j\in S\setminus S^{P}$ and $q^S_j=Q_j$%
. We consider two cases again:

\begin{enumerate}
\item $\sum_{i\in R^{S} }Q_i\leq \underline{m}$: Consider an alternative
solution with $q_{j} =0$ , $q_i=q^S_i$ for all $i\in S\setminus \{j\}$, and $%
m= \underline{m} $. We have $\Pi^S(q,m)- \Pi^S(q^S,m^S)= -\delta_j \geq 0$
which is a contradiction.

\item $\sum_{i\in R^{S}\setminus \{j\}}Q_i\geq \underline{m}$: Consider an
alternative solution with $q_{j} =0$ , $q_i=q^S_i$ for all $i\in S\setminus
\{j\}$, and $m= \sum_{i\in R^{S}\setminus \{j\}}Q_i$. We have $\Pi^S(q,m)-
\Pi^S(q^S,m^S)= -\Delta_j \geq 0$ which is a contradiction.

\item $\sum_{i\in R^{S}\setminus \{j\}}Q_i< \underline{m}$ and $\sum_{i\in R^{S}
}Q_i> \underline{m}$: Consider an alternative solution with $q_{j} =0$, $%
q_i=q^S_i$ for all $i\in S\setminus \{j\}$, and $m=\underline{m}$. We have $%
\Pi^S(q,m)- \Pi^S(q^S,m^S)= -\Delta_j-\left(\underline{m}-\sum_{i\in
R^{S}\setminus \{j\}}Q_i\right)r^u \geq 0$ which is a contradiction.
\end{enumerate}

Thus $S^{E}\subseteq R^{S}\subseteq S^{P}$.
\end{proof}

\begin{proof}[Proof of Corollary \ref{coro0}]
    Considering  Theorem \ref{th3}, in case (A) we have $\Pi(q^S,m^S)=\sum_{i\in S^{E}}\Delta_i$. In case (B) we have
     $\Pi(q^S,m^S)=\sum_{i\in S^{E}}\Delta_i+\sum_{i\in R^{S}\setminus S^{E}}\Delta_i-\left(\underline{m}-\sum_{i\in R^{S}}Q_i\right)^+r^u$. Note that by definition of complementary players, for any $i\in R^{S}\setminus S^{E}$ we have $\Delta_i<0$ which obtains  $\Pi(q^S,m^S)\leq \sum_{i\in S^{E}}\Delta_i $.
\end{proof}

\begin{proof}[Proof of Proposition \ref{lemcompl}]
Take $S \in N$ and suppose suppose that $\sum_{i\in S^{E}}Q_i<\underline{m}$. 
If $R^{S}\neq \emptyset$, we know by Theorem \ref{th3} (B) that $S^{E}\subseteq R^{S}$. Consider $\Pi^S(S^{E})$ as the profit function for coalition $S$ in the solution in which only the essential players export all of their capacity.  Thus we can write 
\begin{align*}
  \Pi^S(S^{E})+G^S(D)=& \sum_{i\in S^{E}}\Delta_i  -\left( \underline{m} -\sum_{i\in S^{E}} Q_{i}\right)^{+}r^{u}   + \sum_{i\in D}\Delta_i+\min\left\{\underline{m}-\sum_{i\in S^{E}}Q_i,\sum_{i\in D}Q_i\right\}r^u \\
=& \sum_{i\in S^{E}}\Delta_i + \sum_{i\in D}\Delta_i  -\left( \underline{m}-\sum_{i\in S^{E}} Q_{i}\right) r^{u} +\min\left\{\underline{m}-\sum_{i\in S^{E}}Q_i,\sum_{i\in D}Q_i\right\}r^u \\
=& \sum_{i\in S^{E}\cup D}\Delta_i  -\left( \underline{m}\sum_{i\in S^{E} \cup D} Q_{i}\right)^{+}r^{u} = \Pi^S(S^{E}\cup D)
\end{align*}
Hence, finding optimal $R^{S}$ boils down to finding optimal $D$ for $G^ S$; that is, $D^S\in \argmax_{D \subseteq S^{C}}G^{S}(D)$.    
\end{proof}

\begin{proof}[Proof of Proposition \ref{NPhard}]
     Consider $D^S$ such that $\max_{D\subseteq S^{C}} \{ G^S(D) \} = G^S(D^S)$. Assume that $D^S$ in non empty. Suppose $D^S\neq \emptyset$ and let $M=\underline{m}-\sum_{i\in S^{E}}Q_i$. 

Suppose an oracle indicates that at optimality, $\sum_{i\in D^S}Q_i\leq M$. 
Thus we  have $G^S(D^S)=\sum_{i\in D^S}\delta_i$. 
In that case we can search among all groups of players whose total quantities does not exceed $M$. This can be formulated through the following integer program:
\begin{align*}
    \max  &\sum_{i\in  S^{C} }x_i\delta_i\\
    \text{s.t.} & \sum_{i\in  S^{C} }x_iQ_i<M\\
    & x_i\in\{0,1\} \quad \quad \forall i\in S^{C}
\end{align*}
Given the optimal solution $x^*$ we have $D^S=\{i:x^*_i=1\}$. However, the program above is the $\{0,1\}$-Knapsack Problem which is NP-hard \citep{pisinger1998knapsack}.   
\end{proof}

\begin{proof}[Proof of Lemma \ref{lemmma}]
(i) From Proposition \ref{lemcompl} we can write 
\begin{equation*}
D^{N}\in \argmax_{D\subseteq N^{C}}\sum_{i\in D}\Delta _{i}+\min \left\{ 
\underline{m}-\sum_{i\in N^{E}}Q_{i},\sum_{i\in D}Q_{i}\right\} r^{u}.
\end{equation*}%
Thus $G^{N}(D^{N})=\sum_{i\in D^{N}}\Delta _{i}+\min \left\{ \underline{m}%
-\sum_{i\in N^{E}}Q_{i},\sum_{i\in D^{N}}Q_{i}\right\} r^{u}.$ In the same
manner we have 
\begin{equation*}
D^{S}\in \argmax_{D\subseteq N^{C}}\sum_{i\in D}\Delta _{i}+\min \left\{ 
\underline{m}-\sum_{i\in S^{E}}Q_{i},\sum_{i\in D}Q_{i}\right\} r^{u}.
\end{equation*}%
Thus, $G^{S}(D^{S})=\sum_{i\in D^{S}}\Delta _{i}+\min \left\{ \underline{m}%
-\sum_{i\in S^{E}}Q_{i},\sum_{i\in D^{S}}Q_{i}\right\} r^{u}.$

Note that $\underline{m}-\sum_{i\in N^{E}}Q_i\leq \underline{m}-\sum_{i\in
S^{E}}Q_i$.

We first show that $G^S(D^S)\geq G^N(D^N)$. To see this, note that if we use 
$D^N$ in $G^S$ we have $G^S(D^S)\geq G^S(D^N)$ by definition of optimality
and considering that the two optimization problems have the same choice set.
Moreover, we have $G^S(D^N)\geq G^N(D^N)$ because 
\begin{equation*}
\min\left\{\underline{m}-\sum_{i\in S^{E}}Q_i,\sum_{i\in D^N}Q_i\right\}\geq
\min\left\{\underline{m}-\sum_{i\in N^{E}}Q_i,\sum_{i\in D^N}Q_i\right\}.
\end{equation*}
Thus, we have 
\begin{equation}  \label{pr1}
\sum_{i\in D^S}\Delta_i+\min\left\{\underline{m}-\sum_{i\in
S^{E}}Q_i,\sum_{i\in D^S}Q_i\right\}r^u\geq \sum_{i\in D^N}\Delta_i+\min\left\{%
\underline{m}-\sum_{i\in N^{E}}Q_i,\sum_{i\in D^N}Q_i\right\}r^u.
\end{equation}

Next, suppose the contrary, that is $\sum_{i\in D^S}\Delta_i> \sum_{i\in
D^N}\Delta_i. $ Consider the value of $G^N$ at $D^S$, that is, 
\begin{equation*}
G^N(D^S)= \sum_{i\in D^S}\Delta_i+\min\left\{\underline{m}-\sum_{i\in
N^{E}}Q_i,\sum_{i\in D^S}Q_i\right\}r^u.
\end{equation*}
We consider two cases:

\noindent [Case 1] Suppose $\sum_{i\in D^S}Q_i\geq \underline{m}-\sum_{i\in {S%
}^{E}}Q_i$: From the fact that $\underline{m}-\sum_{i\in S^{E}}Q_i>\underline{m}%
-\sum_{i\in N^{E}}Q_i$ we get 
\begin{equation*}
G^N(D^S)=\sum_{i\in D^S}\Delta_i+ \left(\underline{m}-\sum_{i\in
N^{E}}Q_i\right)r^u
\end{equation*}
From equation (\ref{pr1}) we get 
\begin{equation*}
G^N(D^S)= \sum_{i\in D^S}\Delta_i+ \left(\underline{m}-\sum_{i\in \overline{N%
}^{E}}Q_i\right)r^u >\sum_{i\in D^N}\Delta_i+\min\left\{\underline{m}%
-\sum_{i\in N^{E}}Q_i,\sum_{i\in D^N}Q_i\right\}r^u =G^N(D^N).
\end{equation*}
However, the above contradicts the optimality $D^N$ for $N$. \newline

\noindent [Case 2] Suppose $\sum_{i\in D^S}Q_i< \underline{m}-\sum_{i\in S^{E}}Q_i$: We
consider two further subcases:

[Case 2.a] If $\underline{m}-\sum_{i\in N^{E}}Q_i\leq \sum_{i\in D^S}Q_i$, we
have 
\begin{equation*}
G^N(D^S)=\sum_{i\in D^S}\Delta_i+ \min\left\{\underline{m}-\sum_{i\in
N^{E}}Q_i, \sum_{i\in D^S}Q_i \right\}r^u\geq \sum_{i\in D^N}\Delta_i+
\min\left\{\underline{m}-\sum_{i\in N^{E}}Q_i, \sum_{i\in D^N}Q_i
\right\}r^u=G^N(D^N).
\end{equation*}
The above, however, is a contradiction.

[Case 2.b] If $\underline{m}-\sum_{i\in N^{E}}Q_i> \sum_{i\in D^S}Q_i$, we
have 
\begin{equation*}
G^N(D^S)=\sum_{i\in D^S}\Delta_i+ \sum_{i\in D^S}Q_i r^u\geq \sum_{i\in
D^N}\Delta_i+ \min\left\{\underline{m}-\sum_{i\in N^{E}}Q_i, \sum_{i\in
D^N}Q_i \right\}r^u=G^N(D^N),
\end{equation*}
where the inequality follows from equation (\ref{pr1}). The above is a
contradiction.

Therefore, it must be that $\sum_{i\in D^{S}}\Delta _{i}\leq \sum_{i\in
D^{N}}\Delta _{i}.$

(ii) The statement is equivalent to 
\begin{equation*}
\sum_{i\in D^{S}}\Delta _{i}-\left( \underline{m}-\sum_{i\in
R^{S}}Q_{i}\right) ^{+}r^{u}\leq \sum_{i\in D^{N}}\Delta _{i}-\left( 
\underline{m}-\sum_{i\in R^{N}}Q_{i}\right) ^{+}r^{u}.
\end{equation*}%
Suppose the contrary that is 
\begin{equation*}
\sum_{i\in D^{S}}\Delta _{i}-\left( \underline{m}-\sum_{i\in
R^{S}}Q_{i}\right) ^{+}r^{u}>\sum_{i\in D^{N}}\Delta _{i}-\left( \underline{m%
}-\sum_{i\in R^{N}}Q_{i}\right) ^{+}r^{u}.
\end{equation*}%
First note that since $R^{S}={S}^{E}\cup D^{S}$ we have 
\begin{equation*}
\sum_{i\in D^{S}}\Delta _{i}-\left( \underline{m}-\sum_{i\in {N}%
^{E}\cup D^{S}}Q_{i}\right) ^{+}r^{u}\geq \sum_{i\in D^{S}}\Delta
_{i}-\left( \underline{m}-\sum_{i\in R^{S}}Q_{i}\right) ^{+}r^{u}.
\end{equation*}

Consider the alternative solution for $N$ with $R={N}^{E}\cup D^{S}$%
. We have 
\begin{align*}
\Pi (R)=& \sum_{i\in R}\Delta _{i}-\left( \underline{m}-\sum_{i\in
R}Q_{i}\right) ^{+}r^{u} \\
=& \sum_{i\in {N}^{E}}\Delta _{i}+\sum_{i\in D^{S}}\Delta
_{i}-\left( \underline{m}-\sum_{i\in {N}^{E}\cup D^{S}}Q_{i}\right)
^{+}r^{u} \\
\geq& \sum_{i\in {N}^{E}}\Delta _{i}+\sum_{i\in D^{S}}\Delta
_{i}-\left( \underline{m}-\sum_{i\in R^{S}}Q_{i}\right) ^{+}r^{u} \\
>& \sum_{i\in {N}^{E}}\Delta _{i}+\sum_{i\in D^{N}}\Delta
_{i}-\left( \underline{m}-\sum_{i\in R^{N}}Q_{i}\right) ^{+}r^{u} \\
=& \Pi (R^{N}),
\end{align*}%
where the inequality follows from the contrary assumption. The above implies
that $R^{N}$ is not the optimal solution for $N$ which is a contradiction.
Thus the statement in the lemma holds.
\end{proof}

\begin{proof}[Proof of Proposition \ref{Lemma game}]
(i) if $D\subseteq N\setminus N^{E}$ then we have that $q_{i}^{D}=0$ for $%
i\in D$ for all $i\in D$. Hence, $v(D)=0$.

(ii) If $D_{S}\subseteq S\setminus S^{P}$ then $v(D_{S})=0$ by (1) and $%
R^{S}=R^{S\setminus D_{S}}$ since players of $D_{S}$ never export. $%
v(S\setminus D_{S})=\sum_{i\in R^{S}}\Delta_i -\left( 
\underline{m}-\sum_{i\in R^{S}}Q_{i}\right) ^{+}r^{u}=v(S).$

(iii)%
\begin{eqnarray*}
v(S)+v(T) &=&\sum_{i\in R^{S}}\Delta_i -\left( \underline{m%
}-\sum_{i\in R^{S}}Q_{i}\right) ^{+}r^{u}+\sum_{i\in R^{T}}
\Delta_i -\left( \underline{m}-\sum_{i\in R^{T}}Q_{i}\right)
^{+}r^{u} \\
&\leq &\sum_{i\in R^{S}\cup R^{T}}\Delta_i -\left( 
\underline{m}-\sum_{i\in R^{S}\cup R^{T}}Q_{i}\right) ^{+}r^{u}\leq v(S\cup
T)
\end{eqnarray*}

since $R^{S\cup T}$ is the optimal exporter coalition for coalition $S\cup
T. $

(iv) Immediately follows from (iv) and the non-negative value of the game $%
(N,v)$.
\end{proof}

\begin{proof}[Proof of Theorem \ref{th4}]
The allocation $\phi (N,v) $ is evidently efficient. We can rewrite $\phi _{i}(N,v)$
for each $i\in N$ as follows: 
\begin{equation*}
\phi _{i}(N,v):=%
\begin{cases}
\Delta_i -\left( \left( \underline{m}-\sum_{i\in
R^{N}}Q_{i}\right) ^{+}r^{u}-\sum_{j\in D^{N}}\Delta_i \right) \frac{\Delta _{i}}{\sum_{j\in N^{E}}\Delta _{j}}
& \text{if }i\in N^{E}\text{ and }R^{N}\neq \emptyset \\ 
0 & \text{otherwise}%
\end{cases}%
\end{equation*}%
Then, we check stability for a coalition $S\subset N$ by distinguishing
three cases:

\begin{itemize}
\item[1] When $\sum_{i\in N^{E}}Q_{i}\geq \underline{m}$, then $R^{N}=N^{E}$%
. We check stability for a coalition $S\subset N$. We must show $\sum_{i\in
S}\phi _{i}(N,v)\geq v(S)$ that is 
\begin{equation*}
\sum_{i\in S^{E}}\Delta_i \geq \sum_{i\in R^{S}}\Delta_i -\left( \underline{m}-\sum_{i\in R^{S}}Q_{i}\right)
^{+}r^{u},
\end{equation*}%
which simplifies into 
\begin{equation*}
\sum_{i\in R^{S}\setminus S^{E}}\Delta _{i}-\left( \underline{m}-\sum_{i\in
R^{S}}Q_{i}\right) ^{+}r^{u}\leq 0.
\end{equation*}%
The above always hold since for any $i\in R^{S}\setminus S^{E}$ we have $%
\Delta _{i}<0$. thus the allocation is in the core.

\item[2.] Suppose $\sum_{i\in N^{E}}Q_{i}<\underline{m}$ and $%
R^{N}=\emptyset .$ Then, $v(N)=0$ and by (iv) in Proposition \ref{Lemma game} we
have that $v(S)=0=\sum_{i\in S}\phi _{i}(N,v)$ for all $S\subseteq N.$

\item[3.] Suppose $\sum_{i\in N^{E}}Q_{i}<\underline{m}$ and $R^{N}\neq
\emptyset .$ We have that, $q_{i}^{N}\neq 0$ for all $i\in R^{N}$ and $%
q_{i}^{N}=0$ otherwise, and therefore $\phi _{i}(N,v)=$ $\Delta_i -\left( \left( \underline{m}-\sum_{i\in
R^{N}}Q_{i}\right) ^{+}r^{u}-\sum_{j\in D^{N}}\Delta_i \right) \frac{\Delta _{i}}{\sum_{j\in N^{E}}\Delta _{j}}$
for all $i\in N^{E}$ and $\phi _{i}=0$ otherwise. We consider two subcases:

\item[3.1] If $R^{S}=\emptyset .$%
\begin{eqnarray*}
\sum_{i\in S}\phi _{i}(N,v) &=&\sum_{i\in S^{E}}\Delta_i
-\left( \left( \underline{m}-\sum_{i\in R^{N}}Q_{i}\right)
^{+}r^{u}-\sum_{j\in D^{N}}\Delta_i \right) 
\frac{\sum_{i\in S^{E}}\Delta _{i}}{\sum_{j\in N^{E}}\Delta _{j}} \\
&\geq &\sum_{i\in S^{E}}\Delta_i -\sum_{i\in S^{E}}\Delta_i =0=v(S)
\end{eqnarray*}%
the first inequality follows from next inequality:%
\begin{eqnarray*}
\sum_{i\in S^{E}}\Delta_i &\geq &\left( \left( \underline{m%
}-\sum_{i\in R^{N}}Q_{i}\right) ^{+}r^{u}-\sum_{j\in D^{N} }\Delta_i \right) \frac{\sum_{i\in S^{E}}\Delta _{i}}{%
\sum_{j\in N^{E}}\Delta _{j}}; \\
\sum_{i\in S^{E}}\Delta_i &\geq &\left( \left( \underline{m%
}-\sum_{i\in R^{N}}Q_{i}\right) ^{+}r^{u}-\sum_{j\in D^{N} }\Delta_i \right) \frac{\sum_{i\in S^{E}}\Delta_i }{\sum_{j\in N^{E}}\Delta_i }; \\
1 &\geq &\left( \left( \underline{m}-\sum_{i\in R^{N}}Q_{i}\right)
^{+}r^{u}-\sum_{j\in D^{N}}\Delta_i \right) 
\frac{1}{\sum_{j\in N^{E}}\Delta_i }; \\
\sum_{j\in N^{E}}\Delta_i &\geq &\left( \underline{m}%
-\sum_{i\in R^{N}}Q_{i}\right) ^{+}r^{u}-\sum_{j\in D^{N} }\Delta_i \\
\sum_{j\in R^{N}}\Delta_i &\geq &\left( \underline{m}%
-\sum_{i\in R^{N}}Q_{i}\right) ^{+}r^{u};
\end{eqnarray*}%
The last inequality is true by definition of $v(N)$ and the face that $v(N)\geq 0$. 

\item[3.2] $R^{S}\neq \emptyset .$ We consider special sub-coalition $S$
such that $S=S^{0}\cup N^{C}$ with $S^{0}\subset N^{E}$. We must show that 
$\sum_{i\in S}\phi _{i}(N,v)\geq \Pi ^{S}(R^{S})$ that is 
\begin{align*}
& \sum_{i\in S^{E}}\Delta_i+\frac{\sum_{i\in S^{E}}\Delta _{i}}{%
\sum_{j\in N^{E}}\Delta _{j}}\left[ \sum_{j\in D^{N}}\Delta
_{j}-\left( \underline{m}-\sum_{i\in R^{N}}Q_{i}\right) ^{+}r^{u}\right]  \\
\geq & \sum_{i\in S^{E}}\Delta_i+\sum_{i\in R^{S}\setminus
S^{E}}\Delta_i-\left( \underline{m}-\sum_{i\in R^{S}}Q_{i}\right)
^{+}r^{u}
\end{align*}%
That is 
\begin{align*}
& \frac{\sum_{i\in S^{E}}\Delta _{i}}{\sum_{j\in N^{E}}\Delta _{j}}\left[
\sum_{j\in D^{N}}\Delta _{j}-\left( \underline{m}-\sum_{i\in
R^{N}}Q_{i}\right) ^{+}r^{u}\right]  \\
\geq & \sum_{i\in R^{S}\setminus S^{E}}\Delta _{i}-\left( \underline{m}%
-\sum_{i\in R^{S}}Q_{i}\right) ^{+}r^{u}
\end{align*}%
Both sides are negative thus we must show that 
\begin{align*}
& \frac{\sum_{i\in S^{E}}\Delta _{i}}{\sum_{j\in N^{E}}\Delta _{j}}\left[
\left( \underline{m}-\sum_{i\in R^{N}}Q_{i}\right) ^{+}r^{u}-\sum_{j\in
D^{N}}\Delta _{j}\right]  \\
& \leq \left( \underline{m}-\sum_{i\in R^{S}}Q_{i}\right)
^{+}r^{u}-\sum_{i\in R^{S}\setminus S^{E}}\Delta _{i}
\end{align*}%
Since $\sum_{i\in S^{E}}\Delta _{i}/\sum_{i\in N^{E}}\Delta _{j}\leq 1$, it
suffices to show that 
\begin{equation*}
\left( \underline{m}-\sum_{i\in R^{N}}Q_{i}\right) ^{+}r^{u}-\sum_{j\in
D^{N}}\Delta _{j}\leq \left( \underline{m}-\sum_{i\in
R^{S}}Q_{i}\right) ^{+}r^{u}-\sum_{i\in R^{S}\setminus S^{E}}\Delta _{i}
\end{equation*}%
The above holds by property (ii) in Lemma \ref{lemmma}.
\end{itemize}

Consider now a coalition $S$. Note that $S^{C}\subset N^{C}$. Take $%
\widehat{S}=S^{E}\cup N^{C}$. Then,$\ \widehat{S}=(S\setminus H_{S})\cup
(N^{C}\setminus S^{C})$ with $H_{S}=S\setminus S^{P}$ and $(S\setminus
H_{S})\cap (N^{C}\setminus S^{C})=\emptyset .$ Moreover, as the game is
superadditive it is satisfied that: 
\begin{eqnarray*}
\sum_{i\in \widehat{S}}\phi _{i}(N,v) &\geq &v(\widehat{S})=v(S^{E}\cup
N^{C}); \\
\sum_{i\in \widehat{S}}\phi _{i}(N,v) &\geq &v\left( (S\setminus H_{S})\cup
(N^{C}\setminus S^{C})\right) \geq v(S\setminus H_{S})+v\left(
N^{C}\setminus S^{C}\right) ; \\
\sum_{i\in S\setminus H_{S}}\phi _{i}(N,v)+\sum_{i\in N^{C}\setminus
S^{C}}\phi _{i}(N,v) &\geq &v(S\setminus H_{S})+v\left( N^{C}\setminus
S^{C}\right) ; \\
\sum_{i\in S\setminus H_{S}}\phi _{i}(N,v) &\geq &v(S\setminus H_{S});\text{
(by (i) in Proposition \ref{Lemma game} and }\sum_{i\in N^{C}\setminus
S^{C}}\phi _{i}(N,v)=0\text{)} \\
\sum_{i\in S}\phi _{i}(N,v) &\geq &v(S);\text{(by (ii) in Proposition \ref{Lemma
game} and }\sum_{i\in H_{S}}\phi _{i}(N,v)=0\text{)}
\end{eqnarray*}%
which completes the proof.
\end{proof}

\begin{proof}[Proof of Lemma \protect\ref{lemexc}]
In the first step, we show that there exists no $j\in N\setminus N^{E}$, $%
j\notin \hat{S}$, such that $\hat{S}\cup \{j\}\subset N$. To see this,
suppose there is $j\in N\setminus N^{E}$, $j\notin S$, and $S\cup
\{j\}\subset N$. We have $e(S\cup \{j\},\phi (N,v))-e(S,\phi (N,v))=-[v(S\cup
\{j\})-v(S)]\leq 0$. Therefore, we have $e(S\cup \{j\},\phi (N,v))\leq
e(S,\phi (N,v))$.

We continue in two cases:

[Case I: $R^{S}= \emptyset$] $S\cap N^{E} \neq \emptyset$. Suppose $j\in N^{E}$
such that $j\notin S$ and $S\cup \{j\}\neq N$.

$\bullet $ If $R^{S\cup \{j\}}=\emptyset $, we have 
\begin{equation*}
\frac{e(S\cup \{j\},\phi (N,v))}{|S^{E}|+1}-\frac{e(S,\phi (N,v))}{|S^{E}|}=%
\frac{\frac{\sum_{i\in S^{E}}\Delta _{i}+\Delta _{j}}{\sum_{i\in
N^{E}}\Delta _{i}}}{|S^{E}|+1}v(N)-\frac{\frac{\sum_{i\in S^{E}}\Delta _{i}}{%
\sum_{i\in N^{E}}\Delta _{i}}}{|S^{E}|}v(N)\leq 0
\end{equation*}%
which is equivalent to 
\begin{equation*}
|S^{E}|(\sum_{i\in S^{E}}\Delta _{i}+\Delta _{j})-(|S^{E}|+1)\sum_{i\in
S^{E}}\Delta _{i}\leq 0
\end{equation*}%
That is 
\begin{equation*}
|S^{E}|\Delta _{j}-\sum_{i\in S^{E}}\Delta _{i}\leq 0
\end{equation*}%
that is 
\begin{equation*}
\Delta _{j}\leq \frac{\sum_{i\in S^{E}}\Delta _{i}}{|S^{E}|}
\end{equation*}

$\bullet $ If $R^{S\cup \{j\}}\neq \emptyset $, we have 
\begin{equation*}
\frac{e(S\cup \{j\},\phi (N,v))}{|S^{E}|+1}-\frac{e(S,\phi (N,v))}{|S^{E}|}=%
\frac{\frac{\sum_{i\in S^{E}}\Delta _{i}+\Delta _{j}}{\sum_{i\in
N^{E}}\Delta _{i}}}{|S^{E}|+1}v(N)-\frac{v(S\cup \{j\})}{|S^{E}|+1}-\frac{%
\frac{\sum_{i\in S^{E}}\Delta _{i}}{\sum_{i\in N^{E}}\Delta _{i}}}{|S^{E}|}%
v(N)\leq 0
\end{equation*}%
We have 
\begin{equation*}
|S^{E}|(\sum_{i\in S^{E}}\Delta _{i}+\Delta _{j})-|S^{E}|v(S\cup
\{j\})-(|S^{E}|+1)\sum_{i\in S^{E}}\Delta _{i}\leq 0
\end{equation*}%
That is 
\begin{equation*}
\Delta _{j}-v(S\cup \{j\})\leq \frac{\sum_{i\in S^{E}}\Delta _{i}}{|S^{E}|}
\end{equation*}

[Case II: $R^{S}\neq \emptyset $] In the next step, we show that there
exists no $j\in N^{E}$, $j\notin \hat{S}$, such that $\hat{S}\cup
\{j\}\subset N$. To see this, note that from proof of Theorem \ref{th4} we
know that when $R^{S}\neq \emptyset $, then 
\begin{align*}
\sum_{i\in S}\phi _{i}(N,v)-v(S)=& -\frac{\sum_{i\in S^{E}}\Delta _{i}}{%
\sum_{j\in N^{E}}\Delta _{j}}\left[ \left( \underline{m}-\sum_{i\in
R^{N}}Q_{i}\right) ^{+}r^{u}-\sum_{j\in D^{N}}\Delta _{j}%
\right]  \\
& +\left( \underline{m}-\sum_{i\in R^{S}}Q_{i}\right) ^{+}r^{u}-\sum_{i\in
R^{S}\setminus S^{E}}\Delta _{i}
\end{align*}%
The first part decreases as more essential players are included (since $%
\sum_{i\in S^{E}}\Delta _{i}$ increases). By property (ii) in Lemma \ref{lemmma}, the
bottom part also decreases when more essential players are included. Hence, $%
\sum_{i\in \hat{S}}\phi _{i}(N,v)-v(\hat{S})\geq \sum_{i\in \hat{S}\cup
\{j\}}\phi _{i}(N,v)-v(\hat{S}\cup \{j\})$ and subsequently, $e(\hat{S}\cup
\{j\})/(|\hat{S}^{E}|+1)<e(\hat{S})/|\hat{S}^{E}|$. Thus, if $R^{\hat{S}%
}\neq \emptyset $, there exists no $j\in N$, $j\notin \hat{S}$, such that $%
\hat{S}\cup \{j\}\subset N$. In this case we have 
\begin{equation*}
\hat{S}\in \argmin_{i\subset N:R^{N\setminus \{i\}}\neq \emptyset }\frac{%
e(N\setminus \{i\},\phi (N,v))}{|(N\setminus \{i\})^{E}|}
\end{equation*}%
for $i^{\ast }\in \argmin_{i\in N}\sum_{j\in N\setminus \{i\}}\phi
_{j}(N,v)-v(N\setminus \{i\})$. Thus for all $i\in N$ we have 
\begin{equation*}
\sum_{j\in N\setminus \{i^{\ast }\}}\phi _{j}(N,v)-v(N\setminus \{i^{\ast
}\})\leq \sum_{j\in N\setminus \{i\}}\phi _{j}(N,v)-v(N\setminus \{i\})
\end{equation*}%
that is 
\begin{equation*}
v(N\setminus \{i\})+\phi _{i}(N,v)\leq v(N\setminus \{i^{\ast }\})+\phi
_{i^{\ast }}(N,v)
\end{equation*}%
that is 
\begin{equation*}
v(N)-v(N\setminus \{i\})-\phi _{i}(N,v)\geq v(N)-v(N\setminus \{i^{\ast
}\})-\phi _{i^{\ast }}(N,v)
\end{equation*}%
This can be written in terms of marginal contributions, that is, $M_{i^{\ast
}}(N,v)-M_{i}(N,v)\leq \phi _{i^{\ast }}(N,v)-\phi _{i}(N,v)$ for all $i\in N$.

In order to find $\hat{S}$ we proceed as follows: We start with $\hat{S}%
=N\setminus N^{E}$ and add $j^{\ast }=\min_{j\in N^{E}}\Delta _{j}$. If $R^{%
\hat{S}}\neq \emptyset $, then by part II above we have to find the
minimizes among the sets $N\setminus \{i\}$. Otherwise, if $R^{\hat{S}%
}=\emptyset $, then $\hat{S}$ can be the minimizes. All other subsets does
not need consideration because if we add more essential players, the value $%
\frac{e(S,\phi (N,v))}{|S^{E}|}$ can only be decreased if $R^{S}\neq \emptyset 
$, at this point, adding more essential players by step II make the value
even smaller which brings us back to the case of $N\setminus \{i\}$.

(1) Find $j^{\ast }=\min_{j\in N^{E}}\Delta _{j}$ and let $S=\{j^{\ast
}\}\cup N\setminus N^{E}$. (2) Find $i^{\ast }\in \argmin_{i\subset
N:R^{N\setminus \{i\}}\neq \emptyset }\frac{e(N\setminus \{i\},\phi (N,v))}{%
|(N\setminus \{i\})^{E}|}$. (3) If $e(S,\phi (N,v))<\frac{e(S^{\prime },\hat{%
\phi})}{|S^{\prime }|}$, then $\hat{S}=\{j^{\ast }\}\cup N\setminus N^{E}$,
otherwise, $\hat{S}=N\setminus \{i^{\ast }\}$.
\end{proof}

\begin{proof}[Proof of Theorem \protect\ref{th5}]
The allocation $\psi ^{\rho }(N,v)$ is evidently efficient. Take $%
S\varsubsetneq N,$ if $S^{E}=\emptyset $ it is trivial that $\sum_{i\in
S}\psi _{i}^{\rho }(N,v)\geq 0=v(S).$ Otherwise, if the allocation is in the
core, we must have $\sum_{i\in S}\psi _{i}^{\rho }(N,v)\geq \sum_{i\in
S^{E}}\phi _{i}(N,v)-|S^{E}|\rho \geq v(S)$. That is $\rho \leq \frac{%
\sum_{i\in S^{E}}\phi _{i}(N,v)-v(S)}{|S^{E}|}$ for all $S\subset N$ which
require $\rho \leq \min_{S\varsubsetneq N:S^{E}\neq \emptyset }\left\{ \frac{%
\sum_{i\in S^{E}}\phi _{i}(N,v)-v(S)}{|S^{E}|}\right\} $.
\end{proof}

\begin{proof}[Proof of Lemma \protect\ref{lem4}]
The maximizer of $r(S,\phi (N,v))$ occurs when all non-essential players are
included. We now examine adding more essential players. Let $S\subset N$ and
suppose $j\in N^{E}$ such that $j\notin S$.

\noindent [Case 1] $R^{S}\neq \emptyset $: In order to have $r(S,\phi (N,v))\geq
r(S\cup \{j\},\phi (N,v))$ we have 
\begin{equation*}
\frac{v(S\cup \{j\})}{\sum_{S^{E}}\Delta _{i}+\Delta _{j}}\geq \frac{v(S)}{%
\sum_{S^{E}}\Delta _{i}}
\end{equation*}%
That is $\left[ v(S\cup \{j\})-v(S)\right] \sum_{S^{E}}\Delta _{i}\geq
\Delta _{j}v(S)$. We have $\sum_{i\in S^{E}}\Delta _{i}\geq v(S)$. Also, we
have $v(S\cup \{j\})-v(S)=\Delta _{j}+(y^{S}-y^{S\cup \{j\}})r^{u}\geq
\Delta _{j}$, thus the inequality holds and adding more essential players
will increase the value of $r(S,\phi (N,v))$.

\noindent [Case 2] $R^{S}=\emptyset $: If $R^{S\cup \{j\}}=\emptyset $, we have $%
r(S,\phi (N,v))=r(S\cup \{j\},\phi (N,v))$. Suppose $R^{S\cup \{j\}}\neq
\emptyset $. In this case we have $r(S,\phi (N,v))=0$, and $r(S\cup \{j\},\phi
(N,v))>0$ which completes the proof.
\end{proof}

\begin{proof}[Proof of Theorem \protect\ref{th6}]
The allocation $\varphi ^{\rho }(N,v)$ is evidently efficient. Take $%
S\varsubsetneq N,$ if $S^{E}=\emptyset $ it is trivial that $%
\sum_{i\in S}\varphi _{i}^{\rho }(N,v)\geq 0=v(S).$ Otherwise, if the
allocation is in the core, we must have $\sum_{i\in S}\varphi _{i}^{\rho
}(N,v)\geq (1-\rho )\sum_{i\in S^{E}}\phi _{i}(N,v)\geq v(S)$. That is $\rho
\leq 1-\frac{v(S)}{\sum_{i\in S^{E}}\phi _{i}(N,v)}=\frac{e(S,\phi (N,v))}{\sum_{i\in S^{E}}\phi _{i}(N,v)}$ for all $S\varsubsetneq N$
with $S^{E}\neq \emptyset ,$ which require $\rho \leq \min_{S\varsubsetneq
N:S^{E}\neq \emptyset }\left\{ \frac{e(S,\phi (N,v))}{\sum_{i\in S^{E}}\phi _{i}(N,v)}%
\right\} $. 
\end{proof}

 \end{document}